\documentclass[journal]{IEEEtran}
\usepackage[tbtags]{amsmath}
\usepackage{amsthm, amssymb, amsfonts, bm}
\usepackage{graphicx}
\usepackage{color}
\usepackage{xcolor}
\usepackage{cite}
\usepackage{cool}
\usepackage{cuted}
\usepackage[textsize=footnotesize]{todonotes}
\newtheorem{lemma}{Lemma}

\newtheorem{theorem}{Theorem}
\newtheorem{corollary}{Corollary}

\newcommand*{\hermconj}{^{\mathsf{H}}}
\begin{document}
\title{A Stochastic Geometry Framework for Performance Analysis of RIS-assisted OFDM Cellular Networks}
\author{Guodong~Sun, Fran\c{c}ois~Baccelli, Ke~Feng, Luis~Uzeda~Garcia, Stefano~Paris
\thanks{The work of G. Sun and F. Baccelli was supported by the CIFRE program, under the grant agreement 2021/0304 to NOKIA Networks France and by the European Research Council project titled NEMO under grant ERC 788851. The work of F. Baccelli was also supported by the Horizon Europe project titled INSTINCT under grant SNS 101139161, and by the French National Agency for Research project titled France 2030 PEPR réseaux du Futur under grant ANR-22-PEFT-0010. }
\thanks{G. Sun, F. Baccelli, K. Feng are with Institut national de recherche en sciences et technologies du num\'erique (INRIA), 2 Rue Simone IFF, 75012 PARIS, and with Department d'information, Ecole Normale Sup\'erieure, 45 Rue d'Ulm, 75005, PARIS.(email: guodong.sun$|$francois.baccelli@ens.fr, ke.feng@inria.fr).}
\thanks{G. Sun, L. Garcia, S. Paris are with Nokia Networks France, 12 Rue Jean Bart, 91300 MASSY. (email: luis.uzeda\_garcia$|$stefano.paris@nokia.com).}

}
\maketitle
\begin{abstract}
The reconfigurable intelligent surface (RIS) technology allows one to engineer spatial diversity in complex cellular networks.
This paper provides a stochastic geometry framework for the system-level performance assessment of RIS-assisted networks.
To account for the inherent randomness in the spatial deployments of base stations (BSs) and RISs, we model the RIS placements as point processes (PPs) conditioned on the associated BSs, which are modeled by a Poisson point process (PPP). 
We assume that the system uses the orthogonal frequency division multiplexing (OFDM) technique to exploit the multipath diversity provided by RISs.
The downlink coverage probability and ergodic rate can be evaluated when RISs operate as batched powerless beamformers.
The resulting analytical expressions provide a general methodology for assessing the impact of a parameterized RIS model on system performance.
These RIS PPs can be adapted based on the deployment strategy.
We focus on modeling the RISs as a Mat\'ern cluster process (MCP), where each RIS cluster is a finite PPP within a ring centered on its associated BS. 
This model connects link-level knowledge to system-level impacts, such as overall interference and the effects of imperfect channel state information (CSI). 
It also evaluates key RIS deployment parameters, including batch size and RIS density.  Furthermore, we analyze a variant of RIS placement in which RISs are deployed around coverage holes to demonstrate the framework's flexibility and applicability.
Numerical evaluations of the analytical expressions and Monte-Carlo simulations jointly validate the proposed analytical approach and provide valuable insights into the design of future RIS-assisted cellular networks.
\end{abstract}

\begin{IEEEkeywords}
Reconfigurable intelligent surfaces, stochastic geometry, point process, Mat\'ern cluster process, orthogonal frequency division multiplexing, diversity combining

\end{IEEEkeywords}
\section{Introduction}

The upcoming sixth-generation (6G) networks aim for higher energy efficiency without adding complexity or energy consumption\cite{
chochliouros2021energy, xu2020understanding}.
To achieve this, the reconfigurable intelligent surface (RIS) technology, leveraging a large number of passive elements, represents a promising candidate for transforming the radio propagation environment into a controllable one with minimal power consumption\cite{sharma2021reconfigurable, alexandropoulos2020reconfigurable, dai2020reconfigurable}.
Similar to the relay technology, RIS can enhance spatial diversity by providing alternative propagation paths for the signal, enabling control over the radio propagation environment~\cite{renzo2019smart, dai2020reconfigurable}.
In \cite{di2020reconfigurable, ye2021spatially}, comparative studies between RIS and relay reveal key advantages of RIS in terms of energy efficiency, spectrum efficiency, and implementation complexity.

Compared to half-duplex relays, an RIS extends the classical channel between a base station (BS) and user equipment (UE) by creating additional paths~\cite{di2020reconfigurable}.
For the RIS-assisted channel, which is time-dispersive since the direct and the RIS-reflected signals may not arrive at the same time due to the difference in the radio wave propagation distance, the intersymbol interference (ISI) becomes a significant issue\cite{lu2023single}. 
The orthogonal frequency division multiplexing (OFDM) modulation technique, introduced into cellular networks since the fourth generation (4G) long-term evolution (LTE)\cite{furht2016long}, can efficiently render the multipath channel ISI-free\cite{tse2005fundamentals}. 
To characterize the channel model for RIS-assisted communication links, the authors in~\cite{shtaiwi2021channel} show that only the direct link (i.e., the link between the BS and the UE) and the controllable reflected link (i.e., the link passing through the RIS) should be explicitly modeled.
The authors in \cite{ozdogan2019intelligent} use optical physics to derive a far-field pathloss model, where the reflective elements of the RIS jointly beamform the signal in a desired direction while the elements individually act as scatterers. 

The challenge of characterizing the system performance leveraging the concept of relaying signals via separated propagation paths based on stochastic geometry can be traced back to the research on relays, e.g., \cite {wang2011general, ding2014wireless}.  
This task remains challenging for the RIS technology. 
Preliminary studies on the system-level performance analysis of RIS-assisted wireless networks show that adding RISs to the network can increase the coverage of wireless networks and the area spectral efficiency (SE)~\cite{kishk2020exploiting,zhu2020stochastic}. 
For instance, the authors in \cite{zhu2020stochastic} study the impact of mounting RISs on obstacles like buildings, street lamps, and traffic lights and derive a combinatorial solution to compute the overall probability of line-of-sight communication. 
The authors in~\cite{di2019reflection} evaluate the probability of successful signal reflection considering the spatial distribution and orientation of RISs using a Boolean line segment model. 
The authors in \cite{ghatak2021deploy} investigate the RIS placement problem to achieve high coverage probability in the street using a one-dimensional (1D) stochastic geometry model. The work concludes that RISs should be placed at street intersections to increase coverage probability.
While these results provide some useful insights for the system-level deployment of RISs, such as the optimal density of BSs and RISs to maximize coverage probability, the impact of several key properties of RISs and the correlation between BS and RIS positions have not been investigated yet.

Furthermore, the authors in \cite{lyu2021hybrid} model both BSs and RISs as homogeneous Poisson point processes (PPPs) and associate the typical UE to its nearest BS and RIS.  
With similar setups, the authors in \cite{zhang2021reconfigurable} investigate RIS-assisted multi-cell non-orthogonal multiple access (NOMA) networks.  
To account for the spatial correlation between RISs and BSs, the authors in~\cite{wang2023performance} use the Gauss-Poisson process to model the BSs and RISs. 
These works \cite{lyu2021hybrid,zhang2021reconfigurable,wang2023performance} approximate the composite signal by a Gamma distribution using the moment matching method conditioned on a specific system layout and then integrate it over all possible layouts. 
Namely, the BS-RIS pair is treated as a whole by abstracting their spatial distribution.
In \cite{deng2024modeling}, the authors model the composition of a direct link and a reflected link as the sum of two exponential distributions, leveraging the property of the Erlang distribution. 
Although this work provides an accurate distribution of the composite signal, it focuses on the case where both the direct and reflected links experience Rayleigh fading, and the spatial randomness still requires iterative integration. 
For the case where a UE is served by more than one RIS, the iterative integration steps become intractable. 
In \cite{ye2021spatially}, the authors point out that performance analysis for networks where a multitude of RISs serve UEs is an open challenge.

\subsection{Motivation and Contribution}

The present paper uses stochastic geometry, a powerful analytical tool for wireless system analysis and design, as demonstrated in  \cite{andrews2011tractable,baccelli2010stochastic, baccelli2020random,andrews2016primer}. 
A key advantage of this approach lies in its ability to characterize performance metrics through the Laplace transform of the aggregated interference-noise, as highlighted in~\cite{andrews2016primer}. 
When RISs are strategically deployed to improve system performance, stochastic geometry requires an analytical framework that can accurately model system metrics by accounting for the aggregated effects of direct and reflected signals, interference, and noise, expressed through their Laplace transforms.
Prior works typically incorporate the reflected signal into the direct link model. 
The composite signal can only be accurately characterized when both links are modeled as Rayleigh distributed, which results in an Erlang distributed signal\cite{deng2024modeling}.
Otherwise, the composite signal is approximated by a Gamma distribution using the moment matching method~\cite{lyu2021hybrid,zhang2021reconfigurable,wang2023performance}.  
The signal merging approach becomes computationally intractable as the number of RISs increases, because the possible spatial layouts grow exponentially with the number of RISs.
Motivated by the above discussion, 
\begin{itemize} 
\item we propose a flexible analytical framework to evaluate cellular networks assisted by randomly deployed RISs. 
The framework supports scenarios where the number of RISs is either random (modeled as a PPP) or fixed (modeled as a binomial point process (BPP)), and it allows for adaptable RIS placement areas to model various deployment strategies. 
We focus on a scenario where RISs are deployed around the BSs and extend to a variant that accounts for alleviating coverage holes. 
The main contribution is a general analytical method for evaluating RIS-assisted networks by characterizing the composite signal and deriving spectral efficiency. 
This enables a system-level assessment of RIS impact. 
\item We introduce a novel approach by \textit{modeling the reflected signals from multiple RISs as a shot noise field}, thereby extending the conventional stochastic geometry framework that models only interference as a shot noise field. 
The impact of the signal-interference-noise competition can be modeled as the difference between two non-negative random variables: one representing the aggregated interference and noise, and the other representing the aggregated reflected signals.
This difference, defined over the entire real line, presents a challenge for coverage probability derivation due to the need to separate its positive and negative components from its Laplace transform. 
We employ a contour integral method to address this separation, enabling the analytical derivation of coverage probability and spectral efficiency. 
Key performance metrics are then readily obtained from this analysis.
\item  This methodology provides a quantitative framework for addressing system-level inquiries regarding randomly deployed RISs. For instance, while RISs can benefit the signals, will they also significantly increase system-level interference? If yes, under what condition can this be negligible?  How does the accuracy of imperfect channel state information (CSI) for RIS links (or the limitation of non-ideal RIS hardware) affect overall system performance? How do the geometric parameters impact the system's performance? Given that strategically deploying RIS will enhance coverage in certain areas, what are the best deployment strategies?  
\end{itemize}

\subsection{Organization}
The paper is organized as follows: 
We define the major components for modeling RIS-assisted networks in Section~\ref{section:framework}, including the random placements of both BSs and RISs and the fading of individual links. 
In Section~\ref{section:MCP}, we discuss signal processing and derive analytical expressions for system performance metrics. 
In Section~\ref{section:extensions}, we explore RIS deployment strategies. We begin with a specific strategy that models RISs as an MCP. We then extend the analysis to various RIS placement strategies, addressing different deployment scenarios.
We present extensive simulations and numerical results in Section~\ref{section:numericals}, and the paper is concluded in Section~\ref{section:conclusion}.
{Table~\ref{tab:notation} summarizes the notation used in this work.
\begin{table}[htp!]
\centering	
\caption{List of Symbols}
	\begin{tabular}{|c|l|}
	\hline 
		Symbol & Description  \\
		\hline        
		$\lambda_{\{\cdot\}}$ &  Density of BSs or  RISs.\\
            \hline 
		$\mathbf{x}_i \in \Phi_{\rm BS}$ & Location of the $i$th BS in the BS set.\\
		\hline
   $i\in \mathcal{I}$ & Index of the BSs. \\
		\hline
		$\mathbf{y}_{i,j} \in  \phi_{i}$  & The $j$th RIS within the RIS cluster of the $i$th BS. \\
            \hline 
            $j \in \mathcal{J}_i$&  Index of RISs associated with the $i$th BS.\\
		\hline
        $\Phi_{\rm RIS}$ &  Point process of RISs. \\ 
        \hline 
             $\theta$, $\mathbf{\Theta}$ & RIS phase shifts (scalar and matrix). \\
        \hline
            $\imath$ & Imaginary unit.\\
        \hline
                    $M$, $m$ & Number of RIS elements and their index. \\ 
            \hline
  $Z[n]$ & Channel impulse response in the time domain. \\
              \hline
        $h\in \mathbb{C}$  & Channel gain for individual links.\\
            \hline 
             $D, R$ & Superscripts denoting direct or reflected links. \\ \hline
$\delta[n]$ & Kronecker delta function. \\
  \hline
        $c$ & Speed of light.\\
        \hline
        $N_c$ & Length of the tapped delay line filter.\\ 
        \hline 
            $T_s$ & Antenna sampling resolution.\\
            \hline 
            $\rho \in \mathbb{C}$ & Small-scale fading. \\
            \hline $g(d)$ & Pathloss for a link of length $d$.\\
            \hline $\alpha$ & Pathloss exponent. \\
            \hline $\beta$ & Average received power at 1m reference distance. \\
            \hline
            $\gamma$ & Power of random fading.\\
            \hline 
             $o$&  Subscript denoting the typical UE association. \\
            \hline
            $M_o$ & Number of RIS elements allocated to the typical UE. \\
            \hline
           $G(\mathbf{x}, \mathbf{y})$  
 & Pathloss of a reflected link to a typical UE (vector).\\
            \hline 
            $\zeta \in \mathbb{C}$ & Product of two i.i.d. $\rho$ variables. \\
            \hline 
            $\eta \in \mathbb{C}$  & Sum of multiple $\zeta$ variables. \\
        \hline
            $\mathbb{E}[\cdot], \mathbb{V}[\cdot]$ & Expectation and Variance Operators. \\
            \hline
            $\mathcal{N}(\mu,\sigma^2)$ & Gaussian random variable with mean and variance. \\
            \hline
            $\chi^2$ & Chi-square distributed random variable. \\
            \hline
            $\vartheta$ &Beamwidth of the RIS-reflected beam.\\ 
               \hline
  $\mathcal{Z}[k]$ & Channel response in the frequency domain. \\
  \hline
             $s$, $\bar{s}$ & Transmitted and received signals in the time domain. \\
             \hline
             $w$ & Additive white Gaussian noise. \\
             \hline
             $\mathfrak{s}$, $\bar{\mathfrak{s}}$ & Signals in the frequency domain.\\
             \hline
             $N_s$ & Symbol block length.  \\ \hline
		$P_0$ & BS transmission power.	\\
		\hline
 $Q_{S}$, $Q_{I}$, $\sigma^2_{w}$ & Power of signal, interference, and noise.\\
         \hline 
            $\hat{h}$  & Derived channel gain for codewords.\\
            \hline 
           $Q_{\epsilon}$ & CSI estimation error power with accuracy $\epsilon$. \\
            \hline
            $\epsilon$&
            Proportionality constant for link-level accuracy.\\ \hline 
 $Q_{c}$ & Interference power from a cell. \\
\hline
$\mathcal{G}(x, y, \psi)$ & Pathloss of a reflected link to a typical UE (scalar).\\
            \hline
            $\mathcal{L}_{f}$ & Unilateral Laplace transform of function $f$. \\
            \hline 
            $\mathcal{B}_{f}$ & Bilateral Laplace transform of function $f$. \\  
             \hline
            $\mathsf{P}_c$ & Coverage probability. \\
            \hline 
            $r$ & Distance from typical UE to BS. \\
            \hline
		$T$ & SINR threshold for coverage. \\ 
            \hline
            $\tau$ & Ergodic rate [nat/Hz/s].
		\\
            \hline
            $\mathbb{D}_{\mathbf{x}}$ & Ring cluster centered at $\mathbf{x}$.  \\
            \hline 
            $R_{\rm in}, R_{\rm out}$ & Inner and outer radii of the RIS cluster. \\
    \hline
  $ \varpi$ & Angle of a wedge-shaped area.\\
  \hline  $\mathbf{g}(\mathbf{x}_o, \mathbf{y}_{o,j}, \mathbf{u})$ & Pathloss function for a UE at $\mathbf{u}$.\\
  \hline
  $C_D$, $C_R$ & Blockage penalties for direct and reflected links.\\
		\hline
	\end{tabular}
	\label{tab:notation}
\end{table}
}

\section{Major Components of Framework}\label{section:framework}

\begin{figure}[t!]
    \centering
    \includegraphics[width=1.0\linewidth]{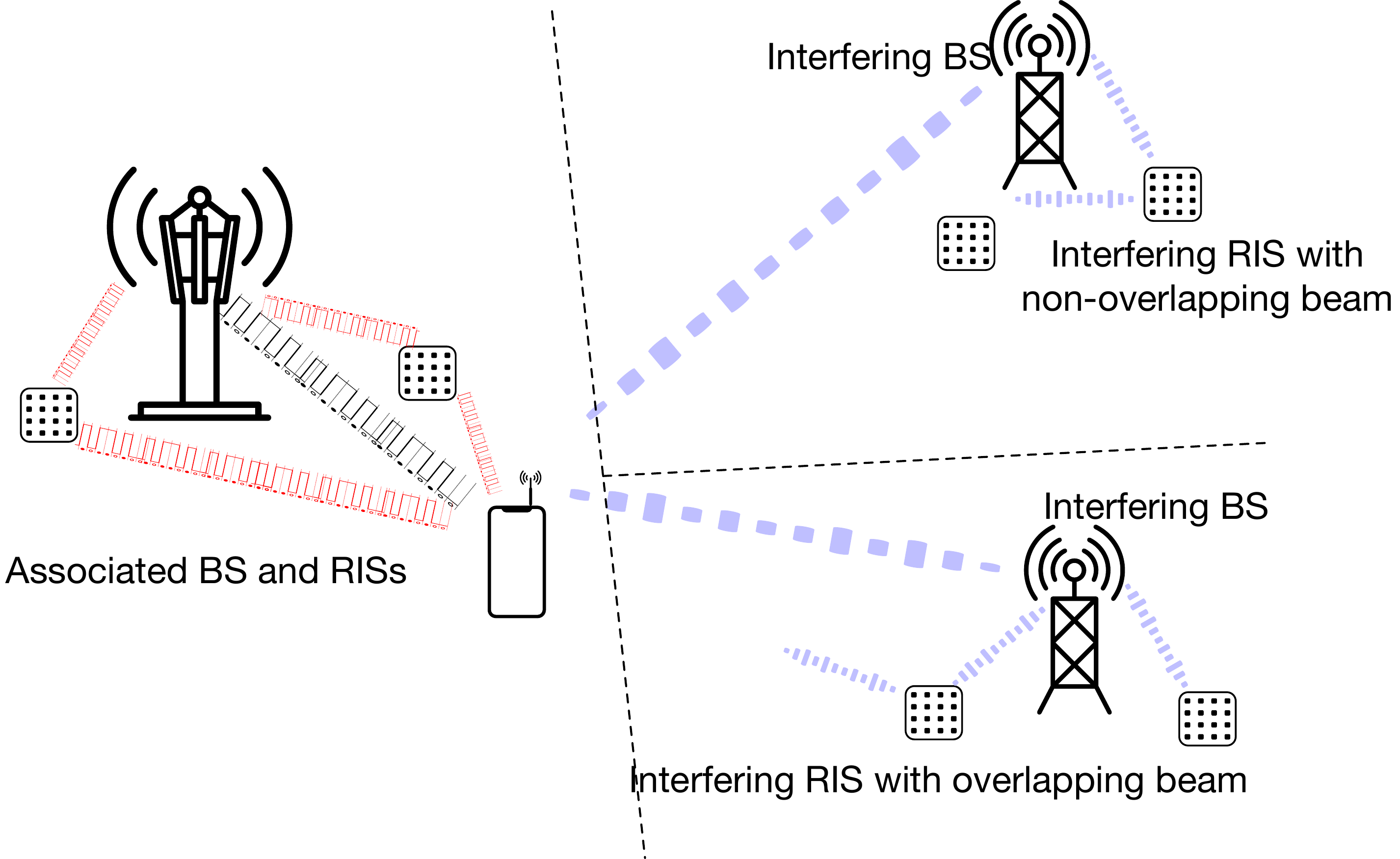}
    \caption{RISs intelligently reflect signals from the associated BSs for their associated UEs, while the BSs and RISs in neighboring cells incur interference.}
    \label{fig:nodes_illustration}
\end{figure}

As illustrated in Fig.~\ref{fig:nodes_illustration}, our model features a cellular network in the two-dimensional (2D) Euclidean space, where randomly located BSs form cells, each containing a random set of RISs. 
We begin this section by modeling these locations as PPs and defining the downlink signal links between BSs, RISs, and a typical UE. 
We then address the fading characteristics of individual links.

\subsection{Stochastic geometry modeling and downlink channels} 

The locations of the BSs are modeled as a homogeneous PPP $ \Phi_{\rm BS} \triangleq \{\mathbf{x}_i\}_{i \in \mathcal{I}}\subset \mathbb{R}^2$ with density $\lambda_{{\rm BS}}$, where $\mathcal{I}$ is the index set of the BSs and each $\mathbf{x}_i$ represents the location of a BS.
Here, BSs transmit signals in all directions to reach randomly located UEs and RISs.
We assume that the UEs are served by their closest BS, and this association policy divides the cellular network into Voronoi cells with respect to (w.r.t.) $\Phi_{\rm BS}$.
Since RISs are network infrastructure entities, we assume that they are managed by their corresponding BS and thus associated with individual cells. 
Specifically, we model the RISs in the Voronoi cell associated with the $i^{\rm th}$ BS by a conditionally independent PP $\phi_i 
 \triangleq \{\mathbf{y}_{i,j}\}_{j \in \mathcal{J}_i}$, where  $\mathbf{y}_{i,j} \in \mathbb{R}^2$ specifies the location of the $j^{\rm th}$ RIS and $\mathcal{J}_i$ is the corresponding index set. 
In other words, the placement of RISs is independent for each cell, only depending on the location of their associated BS $\mathbf{x}_i$.
Then, the overall RISs are modeled as the aggregation of the conditionally independent PPs, $\Phi_{{\rm RIS}}\triangleq  \cup \phi_{i}, i \in \mathcal{I}$.
This RIS PP is intentionally unspecified here to ensure framework flexibility, as the analysis in Section~\ref{section:MCP} is applicable to diverse PPs. 
Specific RIS deployment strategies, defined by the distribution and support of the RIS PP, are detailed in Section~\ref{section:extensions}.

We assume that each RIS, located at $\mathbf{y}_{i,j}$,  has $M$ ideal, lossless reflecting elements, each acting as an independent isotropic phase shifter with a controllable phase shift~\cite{wu2019intelligent, zeng2021reconfigurable}. 
The RIS configuration profile is denoted by the diagonal matrix
\begin{equation}\label{eq:signal_reflection}
\mathbf{\Theta}_{\mathbf{y}_{i,j}} =\text{diag}\{e^{\imath\theta^{(1)}_{\mathbf{y}_{i,j}}}, \ldots, e^{\imath\theta^{(m)}_{\mathbf{y}_{i,j}}}, \ldots, e^{\imath\theta^{(M)}_{\mathbf{y}_{i,j}}} \},
\end{equation} 
where $\theta^{(m)}_{\mathbf{y}_{i,j}} \in [0,2\pi)$ is the phase shift of the $m^{\rm th}$ element. 
Note that we abstract RIS panels as points in our model, neglecting dimensions and deployment orientation. 
We assume that the RISs can effectively serve their associated UEs by reflecting the BS signal into the intended direction according to the RIS-UE association\footnote{RISs can be configured as reflective (serving the UEs on the same side as the BS) or refractive (serving the UEs on the opposite side). Our framework is a common analytical framework, within which specific models of reflective, refractive, or hybrid (reflective and refractive)~\cite{ISG_2023} can be analyzed by specifying the density of the corresponding RIS type.}.

Our analysis focuses on the downlink communication between a BS and a typical UE placed at the origin, assisted by multiple randomly distributed RISs associated with the BS.
These RISs create engineered reflection paths to improve communication.  
The BS signal reaches the UE via both direct and reflected paths with different resolvable delays, making the RIS-assisted channel a time-dispersive wideband channel~\cite{wu2022analysis, lu2023single}.
To model the RIS-assisted channels, we focus on the strongest channel tap for each engineered path  \cite{an2021reconfigurable}, representing each link between a pair of nodes among BS, RIS, and UE as a single channel tap. 
The discrete-time channel impulse response from the $i^{\rm th}$ BS to the typical UE is expressed by
\begin{equation}\label{eq:channel_abbr}
     Z_i[n]  = h_{D_i}\delta[n-n_{D_i}] + \!{\sum_{\mathbf{y}_{i, j} \in \phi_i}\!\!\mathbf{h}_{R_{2,i,j}}\hermconj \mathbf{\Theta}_{{\mathbf{y}_{i, j}}}\mathbf{h}_{R_{1,i,j}}}\delta[n-n_{R,i,j}],  
\end{equation}
where: 1) the channel gains are approximately constant during a transmission block and thus time-independent, denoted by $h_{D_i} \in \mathbb{C}$, $\mathbf{h}_{R_{1, i,j}}\in \mathbb{C}^{M\times 1}$, and $\mathbf{h}_{R_{2, i, j}}\in \mathbb{C}^{M\times 1}$ for the direct link from the $i^{\rm th}$ BS to the typical UE, for the reflected link from the $i^{\rm th}$ BS to $j^{\rm th}$ RIS, and for the link from that RIS to the typical UE, respectively. 
Here, $\{\cdot\}\hermconj$ is the Hermitian conjugate, and the dimension $M$ of $\mathbf{h}_{R_{1, i,j}}$ and $\mathbf{h}_{R_{2, i, j}}$ refers to the number of elements in each RIS;  
2) The discrete-time delays of the direct and the reflected channels are given by $n_{D_i} = \lfloor \frac{\|\mathbf{x}_i\|}{cT_s}\rfloor $ and $n_{R,i,j} = \lfloor \frac{\|\mathbf{x}_i - \mathbf{y}_{i,j}\|+\|\mathbf{y}_{i,j}\|}{cT_s}\rfloor$, respectively.
Here, $c$ is the speed of light, $\frac{1}{T_s}$, is the sampling rate\footnote{This is the time resolution at which the antenna is able to resolve the time delays of the signal traveling via the direct path and the different reflected paths provided by RISs. For example, in the 5G standard, an antenna should support a time resolution of up to 0.509 ns, which allows the antenna to resolve signals passing through different paths with a distance difference of 0.15 meters. This distance is smaller than the distance between most physical nodes, and hence the antenna can resolve different signal paths.}. $\lfloor\cdot \rfloor$ denotes the floor operator, and the abbreviation $[N]$ stands for $\{0, 1, \ldots, N-1\}$. 
The time delay is represented by the Kronecker delta function $\delta[n]$, where $n \in [N_c]$ and $N_c$ is the length of the tapped delay line filter, spanning the channel's delay spread;  
3) We further assume that the resolvable paths have different delays, i.e., no two of the delays $n_{D_i}, n_{R,i,j}$ coincide.
Later, OFDM modulation and demodulation will be applied to mitigate the effect of the time-dispersive channel.
Since OFDM operates on codeword blocks, explicit delay specification is unnecessary. 
In addition, some taps of the sequence $Z_i[n]$ are zeros when there is no significant path at these sampling delays, and this fact is considered in the remaining discussion;
4) We model each RIS as primarily reflecting signals from its associated BS. 
This is justified by the RIS placement, which faces its associated BS and aims to optimize service within its respective cell, rather than for inter-cell communication.   Due to high path loss and the lower probability of line-of-sight (LoS) propagation between cells, inter-cell reflections are negligible compared to the signal from the serving BS.    

For individual links, we assume a fading channel model comprising small-scale fading $\rho \in \mathbb{C}$ and pathloss $g(d)$, where $d$ is the Euclidean distance of the link.
Consequently, the channel gain for the direct link ${h}_{D_i}$ is given by
\begin{equation}\label{eq:fading}
{h}_{D_i} = 
   \rho_{D_i} \sqrt{g(\|\mathbf{x}_i\|)}. 
\end{equation}

Similarly, for the reflected link, the $m^{\rm th}$ entries of the channel gain vectors (marked by superscript)  $\mathbf{h}_{R_{1,i,j}}$ and $\mathbf{h}_{R_{2,i,j}}$ are given by $\rho_{R_{1,i,j}}^{(m)}\sqrt{ g(\|\mathbf{x}_i - \mathbf{y}_{i,j} \|)} $ and $\rho_{R_{2,i,j}}^{(m)} \sqrt{g(\|\mathbf{y}_{i,j}\|)}$, respectively.
The pathloss between two nodes is modeled as $g(d) = \beta(d+1)^{-\alpha}$, where $\alpha>2$ is the path-loss exponent, $\beta = \big(\frac{c}{4\pi f_c}\big)^2$ is the average received power by an isotropic receiving antenna at a reference distance of $1$m based on the free-space path-loss model, and $f_c$ is the carrier frequency. 
We select this path loss model to avoid the singularity of $d^{-\alpha}$ when  $d\rightarrow 0$. 
The multiplicative path loss of each reflected link is given by $G(\mathbf{x}, \mathbf{y}) =g(\|\mathbf{y}\|)g(\|\mathbf{x} - \mathbf{y}\|)$~\cite{ozdogan2019intelligent}, where  $\mathbf{x}\in \Phi_{\rm BS}$ and $\mathbf{y}\in \Phi_{\rm RIS}$ are the BS and RIS locations, respectively. 
The pathloss model, $g(d)$, assumes far-field propagation.  However, if the RIS is not significantly smaller than the RIS-BS/UE distances, near-field effects may compromise the far-field channel model.  Therefore, to ensure consistency with the far-field assumption, our stochastic geometry model employs distance guards to constrain network deployments.

The channel model accounts for two types of multipath phenomena in RIS-assisted networks: the multipath arises from the resolvable propagation distances of the direct and RIS-engineered paths, modeled by time delays in Eq.~\eqref{eq:channel_abbr}; and the multipath causes small-scale fading of individual links due to nearby scatterers, modeled by fading in Eq.~\eqref{eq:fading}.
Reflecting elements of a single RIS share common nearby scatterers and thus share a time-delay window for their fading, which allows for beamforming and one-tap channel modeling.
Environment scattering from distant objects with delays is neglected due to high propagation loss.  

\subsection{Fading power and RIS configuration}

\begin{figure}[t!]
    \centering
    \includegraphics[width=\linewidth]{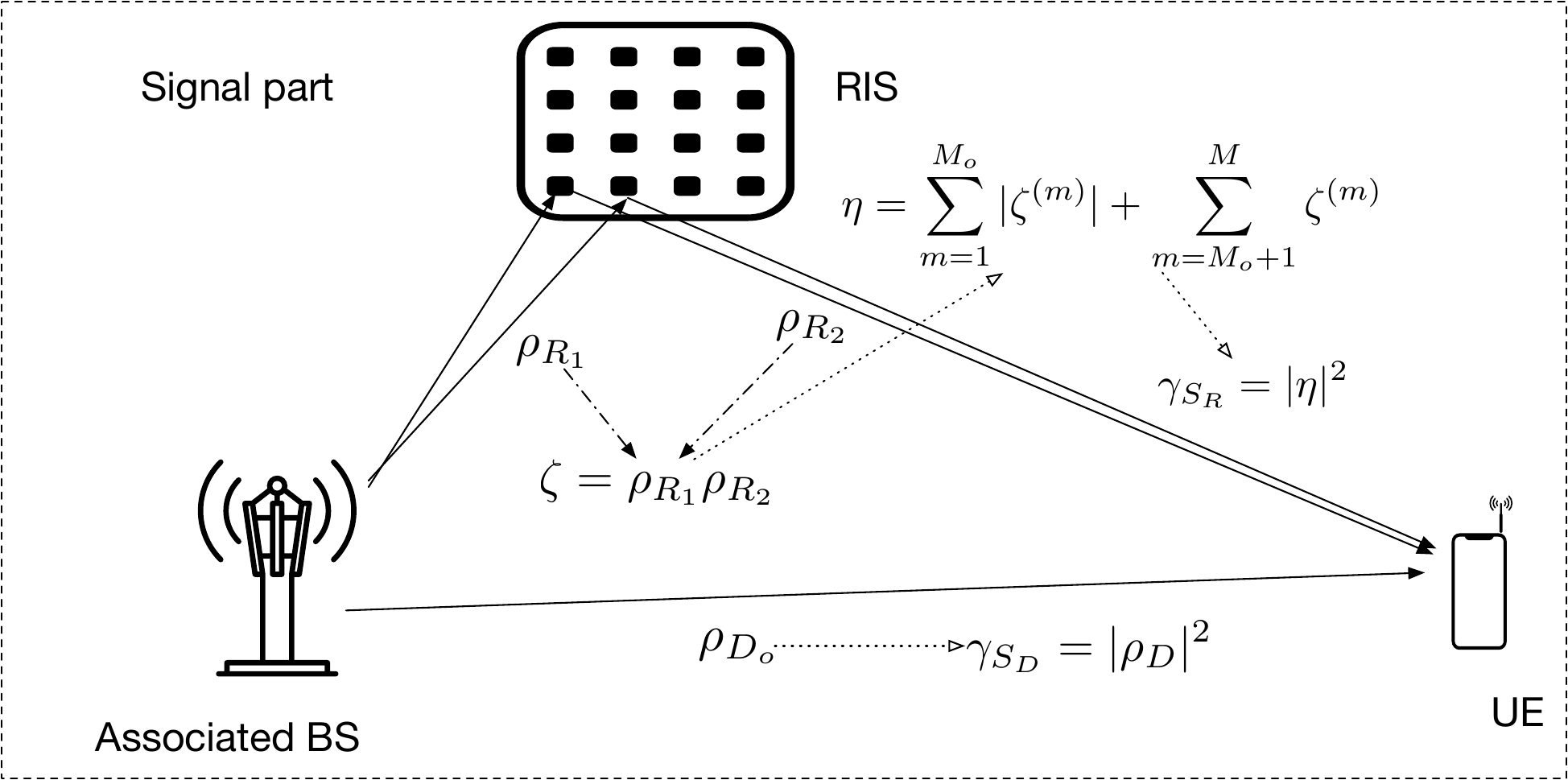}
    \caption{An illustrative overview of the mathematical symbols employed to describe reflection fading and their dependencies}
    \label{fig:SignalFigure}
\end{figure}

For the typical UE, the serving BS $\mathbf{x}_o$ and its corresponding RIS PP $\phi_o$ are indexed by $o$ (as are all associated signals and channels).
Interference from other BSs and RISs indexed by $i \in \mathcal{I}\setminus o$ is taken into account.
We assume orthogonal multiple access within each cell, thus eliminating intra-cell interference. 
We characterize links based on the impact of line-of-sight (LoS) conditions on pathloss exponent and fading power, focusing on the fading power of the direct signal (denoted by  $\gamma_{S_D}$), the reflected signal (denoted by $\gamma_{S_R}$), the direct interference (denoted by  $\gamma_{I_D}$), and the reflected interference (denoted by $\gamma_{I_R}$). 
The relationships among signal propagation, RIS configuration (discussed in detail below), and the consequent power fading are summarized in Fig.~\ref{fig:SignalFigure}.

We model the direct links between BSs and UEs as experiencing non-line-of-sight (NLoS) propagation and the fading $\rho_{D}\in \mathbb{C}$ is modeled as Rayleigh fading, with a path loss exponent $\alpha_{\rm NLoS}$. We have:
\begin{lemma}\label{lemma_signal_fading}
The powers of fading of both the direct signal $|\rho_{D_o}|^2$ and the direct interference $|\rho_{D_i}|^2$ follow a standard exponential distribution\cite{andrews2011tractable}, denoted by $ \gamma_{S_{D_o}}\sim {\rm Exp} (1)$ and $ \gamma_{S_{D_i}}\sim {\rm Exp} (1)$.
\end{lemma}

Given that RISs are strategically deployed to create LoS channels for UEs within their designated cell, we model the channels between the BS and its associated RISs and the channels between these RISs and the associated UE as LoS links. 
The corresponding fading, denoted by  $\rho_{R_{1,o,j}}  \in \mathbb{C}$, $\rho_{R_{1,i,j}}  \in \mathbb{C}$, and $\rho_{R_{2,o,j}}  \in \mathbb{C}$, are modeled as Rician fading with a pathloss exponent $\alpha_{\rm LoS}$. 
On the other hand, the links between the RISs and the UEs in other cells $\rho_{R_{2,i,j}}  \in \mathbb{C}$ are modeled as experiencing NLoS propagation, which will be investigated in Section~\ref{subsec:interference}.
Furthermore, we assume half-wavelength spacing for both transceivers and RISs, so that the fading variables of the direct links $\rho_{D_i}$ and that of the reflected links $\rho_{R_{1,i,j}}^{(m)}$ and $\rho_{R_{2,i,j}}^{(m)}$ can be assumed independent~\cite[Corollary~1]{bjornson2020rayleigh}.
The mutual coupling between RIS elements is neglected in this work.

Reflected signal characteristics depend on RIS configuration.  Since RISs serve multiple UEs at a system level, we assume that each RIS associated with the typical UE allocates a batch of $M_o$ of its $M$ elements for beamforming towards that UE.
Consequently, the RIS configuration profile $\mathbf{\Theta}_{\mathbf{y}_{o,j}}$ in Eq.~\eqref{eq:signal_reflection} is determined batch-wise. 
Defining the phase shifts of the reflected links as the phases of the fadings $\theta_{{R_1}_{o,j}}^{(m)} = \angle \big(\rho_{{R_1}_{o,j}}^{(m)}\big)$ and $\theta_{{R_2}_{o,j}}^{(m)} = \angle \big(\rho_{{R_2}_{o,j}}^{(m)}\big)$, the phase shift configuration for the typical UE's batch is given by:
\begin{equation}
\theta_{\mathbf{y}_{o,j}}^{(m)} = \theta_{{R}_{1, o,j}}^{(1)} +  \theta_{{R}_{2, o,j}}^{( 1)} - \Big(\theta_{{R}_{1, o,j}}^{(m)} +  \theta_{{R}_{2, o,j}}^{(m)}\Big),  1 < m <  M_o, 
\end{equation}
where, without loss of generality, the indices for the typical UE's batch elements start from 1.
This configuration aligns the phases of the reflected signals from the RIS elements of this batch to ensure in-phase arrival at the receiver.
The remaining elements intended for other UEs in the same cell will simultaneously scatter the typical UE signals. 

As shown in Fig.~\ref{fig:SignalFigure}, for a RIS based on the above configuration the small-scale fading of each RIS element is $\zeta = \rho_{R_1}\rho_{R_2} \in \mathbb{C}$, representing the signal variation of the two-link reflection path. 
Its mean and variance are given in Appendix~\ref{app_sub:zeta}.
The reflected signal fading at the typical UE, denoted by $\eta_{o,j} \in \mathbb{C}$, results from the $j^{\rm th}$ RIS configuration, which combines coherent beamforming and non-coherent scattering.
The beamformed component, as the result of in-phase signal fading superposition, is the sum of $M_o$ i.i.d. random variables distributed as $|\zeta|$. 
The scattered component is the sum of the remaining $(M-M_o)$ i.i.d. $\zeta$. 
When specifying the beamformed phase-shift as the reference phase of the received signal fading, $\eta_{o,j}$ is given by 
\begin{equation}\label{eq:chi}
	\eta_{o,j} = \sum_{m=1}^{M_o}\big| \zeta^{(m)} \big| + \sum_{m=M_o+1}^{M}\zeta^{(m)}. 
\end{equation}
Here, since both $M_o$ and $M$ are assumed large, thanks to the central limit theorem, we can approximate $\sum_{m=1}^{M_o}\big| \zeta^{(m)}\big|$ by a one-dimensional non-central Gaussian random variable $ \mathcal{N}\left(M_o\mathbb{E}[|\zeta |], M_o \mathbb{V}[|\zeta|]\right)$ and $\sum_{m=M_o+1}^{M}\zeta^{(m)}$ by a circular complex Gaussian random variable $\mathcal{CN}\left(0, (M-M_o) \mathbb{V}[|\zeta|]\right)$.
A circular Gaussian random variable can be decomposed into two components: one aligned with the beamformed signal phase and another orthogonal to it.
Then, the real part of $\eta_{o,j}$ is expressed as $\Re[\eta_{o,j}] \approx \mathcal{N}\left( \mu, \sigma_{\Re}^2\right)$, where $\mu = M_o\mathbb{E}[|\zeta |]$ and $\sigma_{\Re}^2= \frac{M+M_o}{2}\mathbb{V}[|\zeta|]$.
The imaginary part is  $\Im[\eta_{o,j}] \approx \mathcal{N}\left(0, \sigma_{\Im}^2\right)$, where $\sigma_{\Im}^2 =\frac{M-M_o}{2}\mathbb{V}[|\zeta|]$.

\begin{lemma}\label{lemma_gamma_sr}
The fading power of the reflected signal $\gamma_{S_{R}} = |\eta_{o, j}|^2, \forall j\in\mathcal{J}_{o}$ is the product of two Chi-square distribution, representing the power of the real and imaginary components of $\eta_{o, j}$, respectively.
The corresponding Laplace transform is 
\begin{equation}
    \mathcal{L}_{\gamma_{S_R}}(s) = \frac{ \exp\big(-\frac{\mu^2s}{1+2s\sigma_{\Re}^2}\big)}{\sqrt{(1+2s\sigma_{\Re}^2)(1+2s\sigma_{\Im}^2)}}, \quad s > -\frac{1}{2\sigma_{\Re}^2}.
\end{equation}
\end{lemma}

\begin{proof}
The fading power is the sum of the powers of the real and imaginary parts, modeled as independent non-central and central Gaussian random variables, respectively.  
Since a chi-square distribution $\chi^2$ represents the power of a corresponding normal distribution $\mathcal{N}$,
the Laplace transform of the fading power is the product of the Laplace transforms of these two independent chi-squared random variables, given by
$\mathcal{L}_{\gamma_{S_R}}(s) = \mathcal{L}_{{\chi}^2(\mu, \sigma_{\Re}^2)}(s)\mathcal{L}_{{\chi}^2(0, \sigma_{\Im}^2)}(s)$. The Laplace transform of the chi-squared distribution is given in~\cite{simon2002probability}. The result follows.
\end{proof}

The fading of reflected interference, $\eta_{i,j}$, from a RIS in other cells depends on whether its beam overlaps the typical UE. 
This overlap is probabilistic because the beam is intended for a UE in another cell, not the typical UE.
Let $\vartheta_{\rm beam}$ denote the beamwidth of a reflected beam\footnote{In \cite{han2020half}, the authors demonstrate an inverse relationship between reflected beamwidth and the number of RIS elements, validating a 1-2 degree beamwidth for a 64-element RIS. 
Therefore, with hundreds of elements, the beamwidth can be conservatively approximated as 
$\vartheta_{\rm beam} = \frac{180^\circ}{M_o}$, where $M_o$ is the number of batch elements forming that beam in simulation.}. 
We have:
\begin{lemma}\label{lemma_reflected_interference}
The reflected interference fading power $\gamma_{I_R}$ can be characterized by 
\begin{equation}
    \gamma_{I_R} = \left\{
    \begin{array}{ll}
   \gamma_{S_R},  & \text{with~probability~} \frac{\vartheta_{\rm beam}}{360^\circ}, \\
     {\rm Exp}(M-M_o),& \text{with~probability~} 1- \frac{\vartheta_{\rm beam}}{360^\circ}, 
\end{array}
\right.
\end{equation}
where ${\rm Exp}(M-M_o)$ refers to the exponential distribution with the scale parameter $(M-M_o)$. 
\end{lemma}
\begin{proof}
When an interfering beam covers the typical UE, the UE receives the reflected interference with fading similar to Eq.~\eqref{eq:chi}. Otherwise, the beam is directed elsewhere, and the UE only receives the scattered component, modeled by $\mathcal{CN}\left(0, { (M-M_o) \mathbb{V}[|\zeta|]}\right)$.
The power of this scattered component follows an exponential distribution.
\end{proof}
We use the probabilistic reflected interference fading to characterize total interference incorporating spatial randomness. 
It is important to note that the RIS scattered signal power is often negligible compared to the direct and beamformed reflection signals due to higher multiplicative pathloss~\cite{shtaiwi2021channel}.
Unlike the beamformed component, this high multiplicative pathloss is not compensated by the beamforming mechanism. 
We will validate this quantitatively in the numerical results section.

\section{Signal Processing and Performance Analysis}\label{section:MCP}
This section first discusses signal processing techniques to aggregate signals and interference from spatially distributed nodes.  
Performance metrics are then expressed using Laplace transforms. These metrics capture various random effects, such as spatial distribution and signal variation, and can be adapted to specific RIS deployments.

\subsection{Signal Processing}
We first consider the ideal scenario where the typical UE has perfect CSI $Z_o[n],~n\in[N_c]$. 
This is achieved by the associated BS transmitting pilot signals on a dedicated channel, allowing the UE to perform channel estimation.
The typical UE is agnostic to $Z_i[n],~n\in[N_c]$.
Although perfect CSI is assumed for the initial system-level modeling, the impact of imperfect CSI will be incorporated in a later analysis.

Signals and channels can be analyzed in both the time and frequency domains. This duality means that modulating a signal in one domain will be reflected by its representation in the other domain.
In an OFDM system, we assume the carrier of a total bandwidth $B$ to be equally divided into $N_s$ subcarriers. A codeword of $N_s$ symbols, denoted by $\mathfrak{s}_i[k],~k\in [N_s]$,  is modulated on the BS at $\mathbf{x}_i$ over the OFDM subcarriers. 
This codeword is transmitted as a symbol sequence in the time domain $s_i[n],~n\in [N_s]$, obtained by the inverse discrete Fourier transform of the codeword in the frequency domain
\begin{equation}\label{eq:signal_DFT}
    s_i[n] = \frac{1}{\sqrt{N_s}}\sum_{k=0}^{N_s-1} \mathfrak{s}_i[k] \exp\Big(\frac{\imath 2\pi kn}{N_s}\Big).
\end{equation} 
For the channel $Z_i[n]$, we use $\mathcal{Z}_i[k]$ to denote the channels in the frequency domain. 
Here, we use $n$ and $k$ to index the symbol block in the time and frequency domain, respectively. 

Recall that $N_c$ is the time dispersion duration of the discrete channel. 
In the OFDM system, the symbol block is appended with a cyclic prefix of length $N_c-1$ to mitigate inter-symbol interference (ISI). We further assume that $N_s \gg N_c$, and thus the frequency response of individual subcarriers is flat and the impact of the cyclic prefix is negligible and we neglect it\footnote{Note that the OFDM technique appends the cyclic prefix of length $N_c-1$ to the symbol block consisting of the end of the sequence, and then removes the last $N_c-1$ received symbols at the receiver side. 
As a result, the length of the symbol block in the time domain remains $N_s$. }\cite{tse2005fundamentals}. 
As mentioned earlier, ${s}_o[n]\in \mathbb{C}$ denotes the signal symbol blocks w.r.t. the typical UE. 
In this cellular network, the remaining BSs send signals $s_i[n]$ over the same frequency band, creating system-level interference. 
Let $\overline{s}_o[n], n\in [N_s]$ define the signal received at the typical UE, given by 
\begin{equation}\label{eq:model}
\overline{s}_o[n] = {Z}_o[n] \circledast {s}_o[n] + \sum_{i\neq o} \Big(Z_i[n] \circledast {s}_i[n]\Big) + w[n],
\end{equation}
where $\circledast$ is the circular convolution operation\cite{tse2005fundamentals}.  We recall that $Z_o[n]$ is the channel of the BS to which the typical UE is associated, defined in Eq.~\eqref{eq:channel_abbr}.
The first term in Eq~\eqref{eq:model} will be called signal and the second interference.
Here, $w[n]$ denotes the white Gaussian noise.

Network performance will be evaluated using the Shannon channel capacity, which depends on the signal-to-interference-plus-noise ratio (SINR). 
Specifically, the maximum number of nats per symbol that can be reliably communicated is a function of SINR, given by
 $\log \Big( 1 + \frac{Q_{S}}{Q_{I}+\sigma^2_{w}}\Big),$
where $Q_S$ is the average signal power of the received codeword, whereas $Q_{I}$ and $\sigma^2_{w}$ are the interference and noise power, respectively. 
In this work, we assume that the channel coding is applied jointly across subcarriers, for which the average received power\footnote{In frequency-selective channels, subcarrier gains vary across subcarriers, potentially leading to higher error probabilities for some codewords under average power constraints. 
However, the decoding error depends on the error correction code's efficiency, which is beyond the scope of this work. 
In practice, when the exact CSI is available, the waterfilling algorithm, for example, can optimize power allocation across subcarriers, increasing practical channel capacity beyond the evaluated capacity.} for the codewords depends on the channel gain given in Lemma~\ref{lemma:OFDM_relationship}.


Consider a symbol transmitted over the $k^{\rm th}$ subcarrier, denoted by $\mathfrak{s}_o[k], k\in[N_s]$, the signal power defined in terms of average energy per symbol time is
\begin{equation}
\begin{aligned}
\mathbb{E}\big[\mathfrak{s}_o^2[k]\big]
= \frac{1}{N_s}\mathbb{E}\bigg[ \bigg| \sum_{n=0}^{N_s-1}s_o[n] \exp\bigg(\frac{-\imath 2\pi kn}{N_s}\bigg)\bigg|^2\bigg] \stackrel{(a)}{=}  P_0,
\end{aligned}
\end{equation}
where $(a)$ follows from the fact that, in most modulation schemes, such as QPSK or QAM, the transmitted signals $s_i[n], n\in[N_s]$ are i.i.d. random variables with zero mean and variance $P_0$. 
Note that the transmission of this symbol $\mathfrak{s}_o[k]$ over one OFDM subcarrier of bandwidth $\frac{B}{N_s}$, as shown in Eq.~\eqref{eq:signal_DFT}, takes $N_s$ symbol times.
Hence, the transmission power of each subcarrier is $\frac{P_0}{N_s}$. 

To recover the encoded codeword $\mathfrak{s}_o[k], k\in [N_s]$, the receiver applies the discrete Fourier transform to the received signal and interference $\overline{s}_o[n], k\in [N_s]$ in Eq.~\eqref{eq:model}, expressed as 
\begin{equation}\label{eq:model_freq}
    \overline{\mathfrak{s}}_o[k] = \mathcal{Z}_o[k] \cdot \mathfrak{s}_o[k] + \sum_{i\neq o} \Big(\mathcal{Z}_i[k]  \cdot \mathfrak{s}_i[k]\Big) + w[k]. 
\end{equation}
Here, $\mathcal{Z}_o[k], k\in [N_s]$ denotes the channel gains over subcarriers, given by $\mathcal{Z}_o[k]=\sum_{n=0}^{N_s-1}Z_o[n]\exp\big(\frac{-\imath 2\pi nk}{N_s}\big), k \in [N_s], $
where we append zeros $Z_o[n]=0 $ for all $ n\in \{N_c, \ldots, N_s-1\}$. 
The channel gain over subcarriers is obtained using a scaled DFT, as used in Eq.~\eqref{eq:signal_DFT}, where the scaling factor $\sqrt{N_s}$, ensuring the transformed value equals the channel's frequency response \cite[Eq.~(3.138)]{tse2005fundamentals}\footnote{
The scaling factor $\sqrt{N_s}$ normalizes the energy of the channel use, given that a subcarrier with bandwidth $\frac{1}{N_s}$ bandwidth corresponds to $N_s$ time domain channel uses.}. 
The channel gains for the interference $\mathcal{Z}_i[k], k\in [N_s]$ are defined in the same way.

The received signal over the $k^{\rm th}$ subcarrier is $\mathcal{Z}_o[k] \cdot \mathfrak{s}_o[k]$, with a corresponding power $\big|\mathcal{Z}_o[k]\big|^2\frac{P_0}{N_s}$.
 The average codeword power, $Q_S$ (the SINR numerator), is given by 
\begin{equation}\label{eq:define_h_o}
    Q_S = \sum_{k=0}^{N_s-1} \big|\mathcal{Z}_o[k]\big|^2\frac{P_0}{N_s}:=\hat{h}_o^2P_0,
\end{equation}
where $\hat{h}_o^2=\frac{1}{N_s}\sum_{k=0}^{N_s-1}\big|\mathcal{Z}_o[k]\big|^2$. We have

\begin{lemma}\label{lemma:OFDM_relationship}
Under the foregoing assumptions, the scaling factor caused by the RIS-assisted channel for the OFDM signals is given by 
\begin{equation}\label{eq:ofdm_sum_of_power}
    \hat{h}_o^2 =  \gamma_{S_{D_o}} g(\| \mathbf{x}_o\|) + \sum_{j\in \mathcal{J}_{o}}\gamma_{S_{R_{o,j}}}G(\mathbf{x}_{o, j},\mathbf{y}_{o, j}).
    \end{equation} 
\end{lemma}

\begin{proof}
\begin{equation}
\begin{aligned}
\hat{h}_o^2 & =  \frac{1}{N_s}\sum_{n=0}^{N_s-1}\big|\mathcal{Z}_o[k]\big|^2 \stackrel{(a)}{=}  \sum_{n=0}^{N_s-1}\big|Z_o[n]\big|^2 \\ &\stackrel{(b)}{=} \big| h_{D_o}\delta(n-n_{D_o})\big|^2  +  \sum_{j\in \mathcal{J}_{o}}\big|h_{R_{o,j}}\delta(n-n_{R_{o,j}})\big|^2 \\
&\stackrel{(c)}{=}  \big| h_{D_o} \big|^2  +  \sum_{j\in \mathcal{J}_{o}}\big|h_{R_{o,j}}\big|^2\\
&\stackrel{(d)}{=}\gamma_{S_{D_o}} g(\| \mathbf{x}_o\|) + \sum_{j\in \mathcal{J}_{o}}\gamma_{S_{R_{o,j}}}G(\mathbf{x}_{o, j},\mathbf{y}_{o, j} ),
\end{aligned}
\end{equation}
where $(a)$ holds due to Parseval's theorem\cite{oppenheim1997signals}, the power of the channel gain in the frequency domain equals the power of the multipath taps in the time domain;
$(b)$ from Eq.~\eqref{eq:channel_abbr} holds when the delays of the paths from the BS and the RISs are all different; $(c)$ follows from the fact that the amplitude of the discrete Dirac function is one; $(d)$ follows from the Lemma~\ref{lemma_signal_fading} and Lemma~\ref{lemma_gamma_sr}.
\end{proof}

As a result, OFDM can exploit the multipath diversity gain by mapping the time dispersion of the channel into the frequency domain.

Lemma~\ref{lemma:OFDM_relationship} derives the power of channel gain assuming perfect CSI and the ideal phase shifter model for RISs. 
However, practical scenarios are characterized by imperfect CSI estimation, stemming from limitations such as finite pilot resources, pilot contamination, and specific estimation algorithms. 
We model the CSI error as a Gaussian random variable with power proportional to $Q_{\epsilon}(r)\triangleq \epsilon\mathbb{E}[\hat{h}_o^2]P_0$ consists of the estimation error of all links\cite[Eq.~(8)]{zhang2024channel}. Here, the proportionality constant $\epsilon$ quantifies the desired link-level accuracy. 
The ratio of the channel estimation error power to the transmission power is given by 
\begin{equation}\label{eq:Q_epsilon}
\begin{aligned}
\frac{Q_{\epsilon}(r)}{P_0} =& \epsilon_1\mathbb{E}\big[\gamma_{S_{D_o}}\big] g(\| \mathbf{x}_o\|)\! +\! \epsilon_2  \mathbb{E}\big[\gamma_{S_{R_{o,j}}}\big]\!\sum_{j\in \mathcal{J}_{o}}G(\mathbf{x}_{o, j},\mathbf{y}_{o, j} ) .
\end{aligned}
\end{equation}
where $ \mathbb{E}[\gamma_{S_{D_o}}]=1$ follows from the unit exponential distribution given in Lemma~\ref{lemma_signal_fading} and $\mathbb{E}[\gamma_{S_{R_{o,j}}}] = \mu_{\Re}^2 + \mu_{\Im}^2+\sigma_{\Re}^2+\sigma_{\Im}^2$ follows from the expression for the sum of two Chi-distribution, given in Lemma~\ref{lemma_gamma_sr}.
To account for the difference in maturity between BS-UE and BS-RIS-UE link channel estimation, we utilize different link-level target accuracies, denoted by $\epsilon_1$ and $\epsilon_2$, respectively. 
Although a full link-level analysis of imperfect CSI is beyond the scope of this work, this simplified model allows us to assess its system-level impact, which will be demonstrated via simulation in Section~\ref{subsec:imCSI}.

 
Next, we discuss the fading distribution of the processed interference.
As previously stated, RISs in neighboring cells may randomly reflect beams that overlap the typical UE. 
If beam overlap does not occur, the interference from the RIS is primarily due to scattering.
This probabilistic fading is discussed in Lemma~\ref{lemma_reflected_interference}. 
The same analysis for processing the signal, given in  Eq.~\eqref{eq:define_h_o}, applies to processing the interference, too. 
We define $\hat{h}_{i}^2\in \mathbb{R}$ as the effective channel gain for the interference from the $i^{\rm th}$ interferer, given by  $\hat{h}_{i}^2  \stackrel{}{=} \sum_{n=0}^{N_s-1}\big|Z_i[n]\big|^2$.
Following the same analysis in Lemma~\ref{lemma:OFDM_relationship}, the effective channel gain for the interference from the $i^{\rm th}$ BS is given by
\begin{equation}
        \hat{h}_i^2 =  \gamma_{I_{D_i}} g(\| \mathbf{x}_i\|) + \sum_{j\in \mathcal{J}_{i}}\gamma_{I_{R_{i,j}}}G(\mathbf{x}_{i, j},\mathbf{y}_{i, j} ),
\end{equation}
where both $\gamma_{I_{D_i}}$ and $\gamma_{I_{R_{i,j}}}$ are given in Lemma~\ref{lemma_signal_fading} and Lemma~\ref{lemma_reflected_interference}, respectively. 
The noise term is independent of the BSs and RISs, and we denote the noise power as $\sigma^{2}_{w}$.  

\subsection{Analysis of Coverage Probability and Ergodic Rate}

Next,  we assess the coverage probability and the ergodic rate in an RIS-assisted cellular network.  
The coverage probability is defined as the probability that the SINR of the typical UE is larger than a target threshold $T > 0$. 
Conditioned on the distance from the associated BS to the typical UE $r = \| \mathbf{x}_o \| $, i.e., $\mathsf{P}_{c}(T|r)$, then the coverage probability is
$\mathsf{P}_c(T|r)\triangleq \mathbb{P}\left(\mbox{SINR} \geq T |r\right)$.

In the following, we introduce notations for manipulating expressions. 
To account for the two components of the signal power $Q_{S}(r)$, we let the direct signal power be $Q_{S_D}(r) = \gamma_{S_D} P_0 g(r)$ and the reflected signal power be $Q_{S_R}(r) = \sum_{j\in \mathcal{J}_o} \gamma_{S_{R_{o,j}}} P_0 G(\mathbf{x}_o, \mathbf{y}_{o,j})$.
The total interference power is 
$Q_I = \sum_{i\neq o}  \big( \gamma_{I_{D_i}} P_0 g(\| \mathbf{x}_i\|) + \sum_{j\in \mathcal{J}_i} \gamma_{I_{R_{i,j}}} P_0 G(\mathbf{x}_i, \mathbf{y}_{i,j})\big)$.
Let $\sigma_{w}^2$ denote the power of the Gaussian noise.
Furthermore, we add a term of measurement error accounting for the imperfect CSI $Q_{\epsilon}(r)$\cite[Eq.~(46)]{medard2000effect}.
Then SINR is defined by
\begin{equation}\label{eq:SINR_components_imCSI}
    \text{SINR} = \frac{ Q_{S_D}(r)+Q_{S_R}(r)}{Q_{I}(r)+{ Q_\epsilon(r)} + \sigma_{w}^2 }. 
\end{equation}
Since all the components in Eq.~\eqref{eq:SINR_components_imCSI} and the coverage threshold are non-negative, we can express the coverage probability $\mathsf{P}_{c}(T|r)$ by manipulating the SINR
\begin{equation}\label{eq:cov_expression}
\begin{aligned}
\mathbb{P}\big[Q_{S_D}(r)\geq T(Q_I(r)+\sigma_{w}^2 + {Q_\epsilon(r)})-Q_{S_R}(r)|r\big]. 
\end{aligned}
\end{equation}
Using the introduced notation $Q_{S_D}(r)$ and $Q_{S_R}(r)$, Eq.~\eqref{eq:Q_epsilon} can be expressed as $Q_\epsilon(r) = \epsilon_1 P_0 g(r)+\epsilon_2\mathbb{E}[Q_{S_R}]$.
Next, we denote
\begin{equation}\label{eq:definition}
    \Upsilon = T\big(Q_I(r)+\sigma_{w}^2 {+\epsilon_1 P_0 g(r)+\epsilon_2\mathbb{E}[Q_{S_R}(r)]}\big)-Q_{S_R}(r).
\end{equation} 
Each { independent} component of Eq.~\eqref{eq:definition} has a Laplace transform representation. 

{

\begin{lemma}\label{lemma_1}
The Laplace transform of the distribution function of the aggregated interference power, given that the typical UE is placed at $r = \|\mathbf{x}_o \|$ meters away from its associated BS, is given by
\begin{equation}\label{eq:laplace_interference_BS}
\begin{aligned}
 \mathcal{B}_{Q_{I}(r)}(s) = \exp  \bigg(-2\pi  \lambda_{\rm BS} \int_{r}^{\infty} x \Big(1- \mathcal{L}_{Q_{c}(x)}(s) \Big){\rm d}x\bigg), 
\end{aligned}
\end{equation}
where $Q_{c}(x)$ represents the total interference power from a cell, including both BS and RIS contributions. 
\end{lemma}
\begin{proof}
The total interference power is given by $Q_I(r) = \sum_{i\in \mathcal{I}\setminus o} Q_{c_{i}}$.  By definition, we have $\mathcal{L}_{Q_I(r)}(s) $
\begin{equation}\label{eq:interference_derivation}
\begin{aligned}[b]
\mathcal{L}_{Q_I(r)}(s) &= \mathbb{E}\Big[e^{-sQ_I(r)} \Big|r\Big] = \mathbb{E}_{\Phi_{\rm BS}\setminus \mathbf{x}_o,  Q_{c_{i}} } \Big[ e^{   -s  \sum_{i \neq o} Q_{c_{i}}}\Big] \\
& \stackrel{(a)}{=} \mathbb{E}_{\Phi_{\rm BS}}\Big[ \prod_{i \neq o } \mathbb{E}_{Q_{c_i}} \big[ e^{ -sQ_{c_i}}\big]\Big]\\& \stackrel{(b)}{=} e^{-2\pi \lambda_{\rm BS} \int_{r}^{\infty}  x \big(1-\mathbb{E}_{Q_{c}(x)} \big[ e^{-sQ_{c}(x)}\big] \big)  {\rm d}x} \\
& \stackrel{(c)}{=} e^{-2\pi  \lambda_{\rm BS} \int_{r}^{\infty} x \big(1- \mathcal{L}_{Q_{c}(x)}(s) 
     \big){\rm d}x},
\end{aligned}
\end{equation}
where $(a)$ follows from the i.i.d. distribution of the interference power $Q_{c_i}$ from individual cells and its independence from the PPP $\Phi_{\rm BS}$ defining the random placement of the BSs. 
$(b)$ follows from the probability generating functional (PGFL) of the PPP that $\mathbb{E}[\prod_{x\in \phi}f(x)]\!\stackrel{\phi:{\rm PPP}}{=}\!\exp(-\lambda\int (1\!-\!f(x)){\rm d}x)$, where the integration region is the support of $\phi$. 
The isotropy of the PPP enables the use of polar coordinates in $(b)$, and the integration limits are from $r$ to $\infty$ since the closest BS is at a distance $r$.
Furthermore, the aggregated interference power from the $i^{\rm th}$ cell, $Q_{c_i}$, depends on the distance $x_i$ between the typical UE and the cell's BS.
$(c)$ follows from the definition Laplace transform of the interference power distribution from an arbitrary cell, which will be defined when the RIS deployed is specified. 
\end{proof}


}

The power of both the Gaussian noise $\sigma_{w}^2$ and the BS-UE link's channel estimation error is constant, thus $\mathcal{B}_{\sigma_{w}^2}(sT) = e^{-s\sigma_{w}^2T}$ and $\mathcal{B}_{\epsilon_1g(r)}(sT)=e^{-s\epsilon_1g(r)T}$ are also constants. 
Since $\mathbb{E}[Q_{S_R}(r)]$ and $Q_{S_R}(r)$ are conditioned on the same RIS location PP, they are not independent. Let $\mathcal{B}_{S_R}(s, T)$ denote the Laplace transform of the distribution related to reflected signals $  S_R = \epsilon_2T\mathbb{E}[Q_{S_R}(r)] -  Q_{S_R}(r) $, which will be given in detail in the next section.

Since interference power, noise variance, and estimation error variance are non-negative, the probability distributions of  $Q_I(r)$, $\sigma_{w}^2$, and ${Q_{\epsilon}}$ are defined over $\mathbb{R}^+$.
Similarly, the summed reflected signals $Q_{S_R}(r)$ is defined over $\mathbb{R}^+$ too.  
By definition, the random variable $\Upsilon$ is the difference between the above two non-negative random variables, which is defined over $\mathbb{R}$.
Hence, we use the bilateral Laplace transform\footnote{ 
In functional analysis, the Laplace transform of a function $f$ with positive support usually refers to the unilateral Laplace transform, given by 
$ \mathcal{L}_{f}(s) = \int_{0}^{\infty}f(t)e^{-st}{\rm d}t, \quad t\geq 0$. 
The Laplace transform can be extended to functions with support on the whole real line via the bilateral Laplace transform, given by
$\mathcal{B}_{f}(s) = \int_{-\infty}^{\infty}f(t)e^{-st}{\rm d}t, \quad t\in \mathbb{R}$, under the condition that the integral exists. 
In probability theory, the Laplace transform of a random variable $X$ with density function $f_X$ is defined by $\mathbb{E}[e^{-sX}] = \int_{-\infty}^{\infty} f_X(t)e^{-st}{\rm d}t$, where the range of integration is the support of the random variable $X$. 
Specifically, the Laplace transform is defined by $\mathcal{L}_{X}(s)$ when $X$ is a non-negative random variable and by $\mathcal{B}_{X}(s)$ when $X$ is defined over the entire real axis. 
} $\mathcal{B}_{\Upsilon}(s)$ to characterize the $\Upsilon$, which is obtained by the product of { independent} components\footnote{Notice that for $X$ defined on  $\mathbb{R}^+$, the unilateral Laplace transform $\mathcal{L}_{X}(s)$ and the bilateral Laplace transform $\mathcal{B}_{X}(s)$ are the same.}, 
\begin{equation}\label{eq:cf_upsilon}
\begin{aligned}
\mathcal{B}_{\Upsilon}(s) =&  \mathcal{B}_{\epsilon_1g(r)}(sT)\mathcal{B}_{Q_{I}(r)}(sT)\mathcal{B}_{\sigma_{w}^2}(sT)\mathcal{B}_{S_R}(s, T).
\end{aligned}
\end{equation}

Next, we decompose $\Upsilon = \Upsilon^+ + \Upsilon^-$, where $\Upsilon^+=\max \{0,  \Upsilon\}$ and $  \Upsilon^-=\min \{0,  \Upsilon\}$ to prepare for computing the coverage probability in Lemma~\ref{lemma:coverage}. 
We have the following theorem:
\begin{theorem}
The Laplace transform of the positive part of a random variable can be derived from its bilateral Laplace transform via the formula 
\begin{equation}\label{eq:theorem}
\begin{aligned}
& \mathcal{B}_{\Upsilon^+}(s)= \\ 
& \frac{1}{2\pi \imath}\int_{-\infty}^{\infty} \Big(\mathcal{B}_{\Upsilon}(s-\imath u)- \mathcal{B}_{\Upsilon}(-\imath u) \Big)\frac{{\rm d} u}{u} + \frac{1}{2}\Big( 1+ \mathcal{B}_{\Upsilon}(s)\Big),
\end{aligned}
\end{equation}
where $\int_{-\infty}^{\infty}\frac{{\rm d}u}{u}$ is understood in the sense of Cauchy principal-value, that is $ \int_{-\infty}^{\infty}= \lim_{\epsilon \downarrow 0^+}\int_{-\infty}^{-\epsilon}+  \int_{\epsilon}^{\infty}$.
\end{theorem}
\begin{proof}
    Please refer to Appendix~\ref{app:max_variable}. 
\end{proof}
{ Moreover, the analytical properties of the Laplace transform allows for the differentiation of Eq.~\eqref{eq:theorem} w.r.t. $s$, when the parameter is within the region of convergence\footnote{For a proof of the differentiability properties of the Laplace transform of the positive part of  $\Upsilon$, the reader could refer to our companion work~\cite{sun2025stochastic}, which use this differentiation to analyze performance sensitivity w.r.t. system parameters. }. }
\begin{corollary}\label{coro_derivative}
Applying the Leibniz integral rule to Eq~\eqref{eq:theorem}, we can directly obtain the $k$-th derivatives $\mathcal{B}^{(k)}_{\Upsilon^+}(s)$, 
\begin{equation}\label{eq:derivative}
\mathcal{B}^{(k)}_{\Upsilon^+}(s) = \frac{1}{2\pi \imath} \int_{-\infty}^{\infty} \mathcal{B}_{\Upsilon}^{(k)}(s-\imath u)\frac{{\rm d}u}{u}+\frac{1}{2}\mathcal{B}_{\Upsilon}^{(k)}(s), \quad k>0. 
\end{equation}
\end{corollary}
{
Eqs.~\eqref{eq:theorem} and~\eqref{eq:derivative} provide an analytical representation for the Laplace transform of the positive part of a real-valued random variable. This is particularly useful, as will be shown below.
}

\begin{lemma}\label{lemma:coverage}
When RISs are configured as batched beamformers, the coverage probability for the communication threshold $T$ is given by
\begin{equation}\label{eq:prop1}
\begin{aligned}
\mathsf{P}_c (T|r) =  \mathcal{B}_{ \Upsilon^+ }\left(\frac{1}{P_0g(r)}\right).
\end{aligned}
\end{equation}
\end{lemma}

\begin{proof}
We first compute the coverage probability conditioned on $r$, i.e, $\mathbb{P}(T|r)$ in Eq.~\eqref{eq:cov_expression}, given by
\begin{equation}\label{eq:positive}
	\begin{aligned}[b]
&  \mathbb{P}\left[\gamma_{S_D}  P_0 g(r) \geq \Upsilon\Big|r\right]
=   \mathbb{P}\left[\gamma_{S_D}\geq \frac{\Upsilon}{ P_0 g(r)}\Big| r \right] \\
\stackrel{(a)}{=} & \int_{0}^{\infty} e^{-\frac{\upsilon}{ P_0g(r)}} f_{\Upsilon}(\upsilon){\rm d}\upsilon + \int_{-\infty}^{0}f_{\Upsilon}(\upsilon){\rm d}\upsilon \\
 \stackrel{(b)}{=}&  \mathcal{B}_{ \Upsilon^+ }\left(\frac{1}{P_0g(r)}\right),
\end{aligned}
\end{equation}
where $(a)$ follows from the complementary cumulative distribution function  (CCDF) of the exponential distribution, defined as  $\overline{F}_{{\rm Exp}(1)}(t)=e^{-t}$. Here, the dummy variable $\upsilon$ is integrated over the positive axis since the fading $\gamma_{S_D}$ is defined over the $\mathbb{R}^+$; 
$(b)$ follows from the definition of the Laplace transform of $\mathcal{B}_{\Upsilon^+}(s) = \int_{0}^{\infty}e^{-s\upsilon}f_{\Upsilon}(\upsilon){\rm d}\upsilon + \int_{-\infty}^{0}e^{-s0}f_{\Upsilon}(\upsilon){\rm d}\upsilon$, with $e^{-s0}=1$ for the second integral and the argument as $s=\frac{1}{P_0g(r)}$ .
\end{proof}
Negative-valued $\Upsilon$ implies that coverage is guaranteed by the reflected link, making the coverage probability 1 regardless of the fading of the direct link. 
This contrasts with the coverage probability when $\Upsilon$ takes positive values only.

\paragraph*{Ergodic rate}
Based on the coverage probability $\mathsf{P}_c(T|r)$, we can further obtain the ergodic rate defined by the adaptive Shannon rate, given by~\cite{baccelli2010stochastic}
\begin{equation}\label{eq:ergodic_rate}
\tau(r) \triangleq \mathbb{E}[\log(1+\text{SINR})] = \int_{0}^{\infty} \frac{\mathsf{P}_c(t|r)}{t+1}{\rm d}t.
\end{equation}
The ergodic rate in this context refers to the average performance across random BSs, RIS deployment, random link fading, and interference. However, the distance between the BS and the UE is specified by the fixed parameter $r$. 
This is because several physical constraints will influence the set of UEs that are strategically associated with RISs.

\section{Framework Applications}\label{section:extensions}
The framework design presented in the preceding section allows for direct mapping of the Laplace transforms of the reflected signals and the interference into system performance metrics.  
This section derives these Laplace transforms based on specific RIS deployment strategies.
We focus on a BS-centric model using the MCP to represent RISs and then explore several variants.

{\subsection{MCP model}

\begin{figure}
\centering
\begin{minipage}{.475\linewidth}
  \centering
  \includegraphics[width=\linewidth]{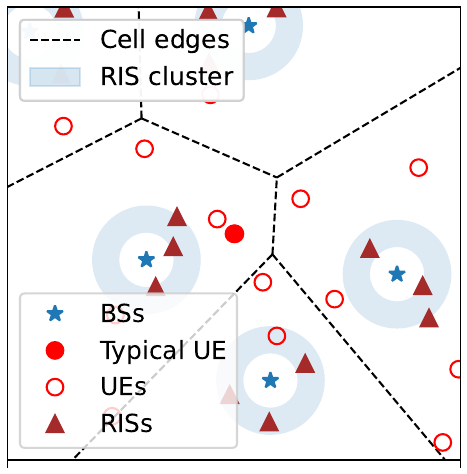}
  \caption{MCP model of a RIS-assisted cellular network (Section IV-A)}
  \label{fig:MCP_model}
\end{minipage}
\hfill
\begin{minipage}{.48\linewidth}
  \centering
  \includegraphics[width=\linewidth]{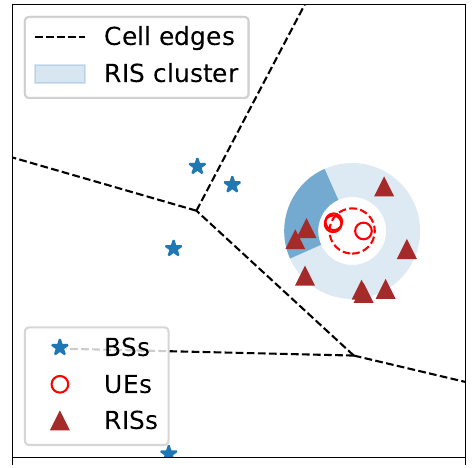}
  \caption{RISs deployed to serve UEs in the coverage hole (Section IV-B)}
  \label{fig:UE_centered_model}
\end{minipage}
\end{figure}

Motivated by the assumption that RISs are directly managed by their associated BSs and are spatially proximate, this BS-centric model exhibits a ring deployment pattern when considering ergodic system-level performance.
This pattern emerges due to spatial averaging across numerous cells, despite the irregular RIS deployment within individual cells.
The ring's radii are determined by two spatial constraints. First, due to the physical limitations of both BSs and RISs, we introduce a minimum separation distance $R_{\rm in}$ between the BS and the RISs. 
This constraint prevents the BS and RISs from being too close, which would invalidate the far-field channel model. 
We also define a maximal separation distance $R_{\rm out}$ between the BS and the RISs, ensuring that the RISs receive and reflect sufficient energy from the associated BS.

The locations of BSs and RISs are thus modeled as an MCP consisting of a \textit{mother point process} and a \textit{daughter point process}, as shown in Fig.~\ref{fig:MCP_model}.
The mother point process of the MCP coincides with the locations of BSs that are modeled as the homogeneous PPP $\Phi_{\rm BS}$. 
The \textit{daughter point process}, conditioned on $\Phi_{\rm BS}$, are PPPs with the density $\lambda_{\rm RIS}$. 
Each cluster $\phi_{i}$ has support on the ring  $\mathbb{D}_{\mathbf{x}_i}(R_{\rm in}, R_{\rm out})$, i.e., the ring centered at $\mathbf{x}_i$ with an inner radius $R_{\rm in}$ and an outer radius $R_{\rm out}$. 
More precisely, the total number of points of $\phi_{i}$ is Poisson distributed with parameter $\lambda_{\rm RIS} \pi (R_{\rm out}^2 -  R_{\rm in}^2) $ and every point of $\phi_{i}$ follows the probability density function (PDF), 
\begin{equation}
\label{eq:mcp_density}
	f_{\text{MCP}}(\mathbf{y}) = \left\{
	    \begin{array}{ll}
        \frac{1}{\pi (R_{\rm out}^2 -  R_{\rm in}^2)} &  \mbox{if } \mathbf{y} \in \mathbb{D}_{\mathbf{x}_i}(R_{\rm out}, R_{\rm in}) , \\
        0 & \mbox{otherwise.}
    \end{array}
	\right.
\end{equation}
Note that this model assumes that RISs are positioned closer to the associated BS than non-associated ones. To ensure this, we should restrict the BS density within a range where the inter-BS distance is significantly larger than twice the outer radius of the clusters. We take here $2R_{\rm out}\ll \frac{1}{2\sqrt{\lambda_{\rm BS}}}$. In RIS-assisted cellular networks that do not satisfy this restriction, signal reflections from non-associated clusters may become significant and must be incorporated into the model.

Having modeled the RISs with the MCP, we can now specify $\mathcal{L}_{Q_c(x)}(s)$ (Eq.~\eqref{eq:laplace_interference_BS}) and $\mathcal{B}_{S_R}(s, T)$ (Eq.~\eqref{eq:cf_upsilon}). 
The reflected signals $S_R$ are subject to two types of randomness: (1) the channel fading power $\gamma_{R}$, and (2) the spatial distribution $\phi_{o}$ of the RISs. 
The aggregated signal power depends on both, whereas the average channel estimation error depends only on $\phi_{o}$. 
In the MCP model, $\phi_{o}$ is a PPP with density $\lambda_{\rm RIS}$ defined on the ring $\mathbb{D}_{\mathbf{x}_o }(R_{\rm in}, R_{\rm out})$. We have 
\begin{lemma}\label{statement1}
The Laplace transform of the distribution of the aggregated reflected signal power, under imperfect CSI and with the typical UE at distance $r = \|\mathbf{x}_o \|$ from its BS, is given by 
\begin{equation}\label{eq:reflecting_signal}
\begin{aligned}
   \mathcal{B}&_{{S_R}(r)}(s, T) =  
 \exp\bigg(-\lambda_{\rm RIS}  \int_{R_{\rm in}}^{R_{\rm out}} \int_{0}^{2\pi}y\Big(1 - \\
    & \mathcal{L}_{\gamma_{S_R} }\big(-s P_0\mathcal{G}(r, y, \psi)\big) e^{-s\epsilon_2T\mathbb{E}[\gamma_{S_{R}}] P_0\mathcal{G}(r, y, \psi)} \Big) {\rm d}\psi {\rm d}y \bigg),
\end{aligned}
\end{equation}
where we use the notation $\mathcal{G}(x, y, \psi) = g(y)g(\sqrt{x^2+y^2-2xy\cos\psi})$ for $ x, y, \in \mathbb{R}, ~ \psi \in [0, 2\pi)$, 
where  $\| \mathbf{y}\| = y$ and $\|\mathbf{x} - \mathbf{y}\| = \sqrt{x^2+y^2-2xy\cos\psi}$, and $\psi$ is the angle between the link from the BS to the UE and that from the BS to the RIS.
\end{lemma}
\begin{proof}
\begin{equation}
\begin{aligned}
&\mathcal{B}_{{S_R}(r)}(s, T)    \\
=& \mathbb{E}_{\phi_{o}, \gamma_{S_{R_{o,j}}}}\bigg[e^{ {\sum_{\mathbf{y}_{o,j}\in \phi_o}  s(\gamma_{S_{R_{o,j}}}-\epsilon_2T\mathbb{E}[\gamma_{S_{R_{o,j}}}]) P_0  G(\mathbf{x}_o, \mathbf{y}_{o,j}) } }\bigg] \\
\stackrel{(a)}{=}& \mathbb{E}_{\phi_{o}}\bigg[\prod_{j=1}^{|\phi_o|} \mathbb{E}_{\gamma_{S_{R_{o,j}}}}\Big[ e^{s(\gamma_{S_{R_{o,j}}}-\epsilon_2T\mathbb{E}[\gamma_{S_{R}}]) P_0  G(\mathbf{x}_o, \mathbf{y}_{o,j}) }\Big]\bigg] \\
=& \mathbb{E}_{\phi_{o}}\!\Big[\!\prod_{j=1}^{|\phi_o|}\! \mathcal{L}_{\gamma_{S_R} }\!\big(\!-\!s P_0 G(\mathbf{x}_o, \mathbf{y}_{o,j}) \big) e^{-s\epsilon_2T\mathbb{E}[\gamma_{S_{R}}] P_0 G(\mathbf{x}_o, \mathbf{y}_{o,j}) }\Big].
\end{aligned}
\end{equation}
where $(a)$ follows from the statistically independent distribution $\gamma_{S_{R_{o,j}}}$ for different RIS links.
Then, Eq.~\eqref{eq:reflecting_signal} follows from the PGFL of PPP $\phi_o$. 
Specifically, we discuss the region of convergence w.r.t. $s$ in Appendix~\ref{app:convergence}. 
\end{proof}

Recall that Eq.~\eqref{eq:laplace_interference_BS} expresses the Laplace transform for the total interference power, $\mathcal{B}_{Q_I(r)}(s)$, in terms of the Laplace transform of the interference power from a single cell $\mathcal{L}_{Q_{c}(x)}(s)$. We have 
\begin{lemma}
    The Laplace transform of the distribution of interference power originating from an arbitrary cell, given that its BS is at a distance $x$, $\mathcal{L}_{Q_{c}(x)}(s)$, is given by
\begin{equation}
\begin{aligned}[b]
    &\mathcal{L}_{Q_{c}(x)}(s)= \frac{1}{1+sP_0g(x_i)}\\
    &e^{-\frac{\vartheta_{\rm beam}}{2\pi}  \lambda_{\rm RIS} \int_{0}^{2\pi}\int_{R_{\rm in}}^{R_{\rm out}} \big(1-\mathcal{L}_{\gamma_{S_R}}(sP_0\mathcal{G}(x_i, y, \psi))\big) {\rm d}y{\rm d}\psi}\\
    &e^{-\big(1-\frac{\vartheta_{\rm beam}}{2\pi}\big) \lambda_{\rm RIS}  \int_{0}^{2\pi}\int_{R_{\rm in}}^{R_{\rm out}} \big(1-\frac{1}{1+s(M-M_o)P_0\mathcal{G}(x_i, y, \psi)}\big) {\rm d}y{\rm d}\psi}. 
\end{aligned}
\end{equation}
\end{lemma}
\begin{proof}
    See Appendix~\ref{app:interference}.
\end{proof}
}

\subsection{Variants}
Different RIS deployment strategies and RIS reflection fadings can be easily adapted in this subsection.
To mitigate coverage holes created by signal blockage from urban buildings, we investigate the deployment of multiple RISs. 
Specifically, we consider a coverage hole centered at a distance $r$ from the BS, and deploy a set of RISs near the hole to enhance the signal coverage for UEs within it. 
The RISs then reflect signals towards those UEs.
Due to the irregular coverage hole shapes and RIS deployments within individual cells, averaging across numerous cells results in a circular pattern, as shown in Fig.~\ref{fig:UE_centered_model}. 
The area of RISs for the coverage hole mitigation scenario is characterized by $\mathbb{D}_{r}(R_{\rm in}, R_{\rm out})$.
Let a circle of radius $R_{\rm CH}<R_{\rm in}$ model the coverage hole within which the typical UE, located at $\mathbf{u}$, can be selected. 
This selection ensures that the UEs are located in the far-field region of the RISs.
We will then discuss how this example can extend the MCP model in three directions.

\subsubsection{Variants of deployment area}

Analyzing this UE-centric model is efficient due to its symmetry with the BS-centric model investigated in the previous section. 
Specifically, the signal propagation paths, BS-RIS-UE in the BS-centric model and UE-RIS-BS in the UE-centric model are essentially mirrored. 
This allows one to directly apply the analysis developed for the BS-centric model, with one adjustment on distance calculations $\mathcal{G}(r, y, \psi)$ in Lemma~\ref{statement1}. 
The new distance calculations will account for the two segments of the reflected path in the new setup: BS-RIS and RIS-UE, given by 
    $\mathbf{g}(\mathbf{x}_o, \mathbf{y}_{o,j}, \mathbf{u}) = g(\|\mathbf{x}_o -  \mathbf{y}_{o,j}\|)g(\|\mathbf{y}_{o,j} - \mathbf{u}\|).$
To further optimize signal coverage, we can extend the deployment beyond a ring shape. 
This allows us to prioritize signal assistance for UEs in specific areas around the coverage hole. 
To model this variant, we consider deploying RIS in a wedge-shaped area with an angle spanning a sector of the ring $\psi \in [-\frac{\varpi}{2}, \frac{\varpi}{2}]$, where $\varpi$ defines the angle of the wedge sector, determined by the environment constraint. 
This modeling requires one to integrate over the area defining the wedge instead of the ring-shape in Lemma~\ref{statement1}, which will be later materialized in Eq.~\eqref{eq:BPP_laplace_transform}. 

Remark that the possible shape of RIS support is flexible too. 
For instance, in a street canyon scenario, a narrow vertical strip deployment of RIS units would be most effective. 
Alternatively, for deploying RISs on top of buildings, a rectangular RIS deployment on the rooftop can be more practical.
For complex deployments, multiple RIS clusters can be defined within a single cell to address intricate coverage needs. 
This flexibility extends beyond 2D modeling, allowing researchers to model linear structures like streets (1D) or incorporate building height for 3D skyscraper scenarios.

\subsubsection{Variants of PP}
Different PPs can be used to model the locations of RISs. 
For example, we can consider models by specifying the exact number of RISs in the area. 
This can be modeled using a BPP, where $N_{\rm RIS} \in \mathbb{N}^+$ represents the number of deployed RISs.
Recall that Lemma~\ref{statement1} characterizes the aggregated reflected signal using the Laplace transform of PPP. 
To incorporate the BPP model defined on the wedge-shaped area, the Laplace transform of the reflected signal power from the associated RISs is given by
\begin{equation}\label{eq:BPP_laplace_transform}
\begin{aligned}
    &\mathcal{B}_{{S_R}(r)}(s, T) =\bigg(\int_{R_{\rm in}}^{R_{\rm out}} \int_{-\frac{\varpi}{2}}^{\frac{\varpi}{2}} \\
    &\mathcal{L}_{\!\gamma_{S_R} }\!\Big(\!\!-\!\!s P_0\mathbf{g}(\mathbf{x}_o, \!\mathbf{y}_{o,j}, \!\mathbf{u}) e^{\!-\!s\epsilon_2T\mathbb{E}[\gamma_{S_{R}}] P_0\mathbf{g}(\mathbf{x}_o, \mathbf{y}_{o,j}, \!\mathbf{u})}\!\Big)
      {\rm d}\psi {\rm d}y\!  \bigg)^{N_{\rm RIS}}_. 
\end{aligned}
\end{equation}
Compared to Equation~\eqref{eq:reflecting_signal}, Equation~\eqref{eq:BPP_laplace_transform} gives the Laplace transform of the combined reflected signals for BPP. 
We observe that the only modifications to accommodate this variant consist in replacing the Laplace transform of the RIS PPP with that of BPP, and updating the integration area. 
Remark that deploying a single RIS can be regarded as a special case in the BPP model by setting $N_{\rm RIS}=1$.

\subsubsection{Variants of blockage}

When the links are severely blocked, some penalty coefficients $C_{D}$  for direct links and $C_{R}$ for reflected links are incurred. 
Then the SINR is replaced by 
\begin{equation}\label{eq:coverage_hole_SINR}
    \frac{ C_{D}Q_{S_D}(r)+ C_{R}Q_{S_R}(r)}{Q_{I}(r)+Q_{\epsilon}(r)+\sigma_{w}^2} \geq T. 
\end{equation}
Using the same manipulation as in Eq.~\eqref{eq:cov_expression}, the coverage probability can be expressed by $\mathbb{P}\big[C_{D}Q_{S_D}(r)\geq T(Q_{I}(r)+Q_{\epsilon}(r)+\sigma_{w}^2) - C_{R}Q_{S_R}(r)\big]$.
Then the coverage probability can be evaluated in terms of the Laplace transform of $\frac{1}{C_{D}}\Upsilon(C_{R}) = \frac{1}{C_{D}}\Big( T(Q_{I}(r)+Q_{\epsilon}(r)+\sigma_{w}^2) - C_{R}Q_{S_R}(r)\Big)$, which can further be written as the Laplace transform of $Q_{I}(r)$ and $Q_{S_R}(r)$. Following the scaling rule of Laplace transforms $f_{\Upsilon}(at) \leftrightarrow \frac{1}{a}\mathcal{B}(\frac{s}{a})$ 
\cite{widder2015laplace}, this evaluation can be easily updated.

It is important to note that this simplified blockage model serves as a foundation for exploring the role of RIS in obstructed environments.

\subsubsection{Scenarios for future exploration of more use cases}
To account for different channel conditions and antenna technologies, fading distribution can be refined given the specification.
For example, replacing the Laplace transform of $\mathcal{L}_{\gamma_{S_R}}(s)$ in Lemma~\ref{statement1} can generalize the fading model to describe variants of the reflected link or different RIS configuration.
Replacing the Laplace transform of $\gamma_{I_{D}}$ and $\gamma_{I_{R}}$ can further generalize the interference characteristics to cope with different scenarios and environment schemes. 
Moreover, extending the direct link beyond Rayleigh fading, for example, Nakagami-$m$ fading, can be adapted by evaluating higher order derivatives of the Laplace transform given by Corollary~\ref{coro_derivative} (see e.g.,\cite{lyu2021hybrid}). 
Remark that all the above extensions discussed in this section are modular and can be performed independently for specific use cases. Our framework thus offers solutions for many more scenarios.

\section{Numerical Results}\label{section:numericals}
This section presents results from both Monte Carlo simulation and analytical calculation\footnote{The numerical simulation code for this work is available on GitHub: https://github.com/Sun8Guden/RISassistedCellularFramework in the following path: /ResearchApplication/RISPerformanceResearch.}. 
To understand the RIS-assisted cellular networks, we first study the overall interference behavior and the effects of imperfect CSI based on the MCP model and further analyze the impact of the MCP parameters. 
Then, we evaluate the extended variant using the BPP model and explore customizable deployment areas and blockage penalties.

\subsection{MCP model:}
{To establish the simulation of the MCP model, we configure each cluster to have an average of 5 RIS panels.
For each cluster, we set the inner radius to 10 meters and the outer radius to 25 meters.
We assume each RIS panel has $M=3000$ elements, and each RIS is serving 5 UEs. In other words, each UE is served by a batch of $M_o=600$ RIS elements.  
The BS density is set as $10/{\rm km}^2$ and we consider the typical UE to be served at a distance $r=100$m. 
We assume an average BS transmission power of $P_0=1$ Watt
and a Gaussian noise power of $\sigma^2_{w}=10^{-13}$ Watt for the environment.
The pathloss exponents are set to $\alpha_{\rm NLoS} = 4$ for NLoS channels and $\alpha_{\rm LoS} = 3$ for LoS channels in general, but the pathloss exponent for the reflected link across cells is specified as $\alpha_{I_R} = \{3, 3.5, 4\}$ for further investigation.
With the exception of Section~\ref{subsec:imCSI}, we operate under the assumption of perfect CSI, characterized by $\epsilon_1=\epsilon_2=0$.
}

\subsubsection{Impact of RIS reflection on interference}\label{subsec:interference}
\begin{figure}
    \centering
    \includegraphics[width=0.8\linewidth]{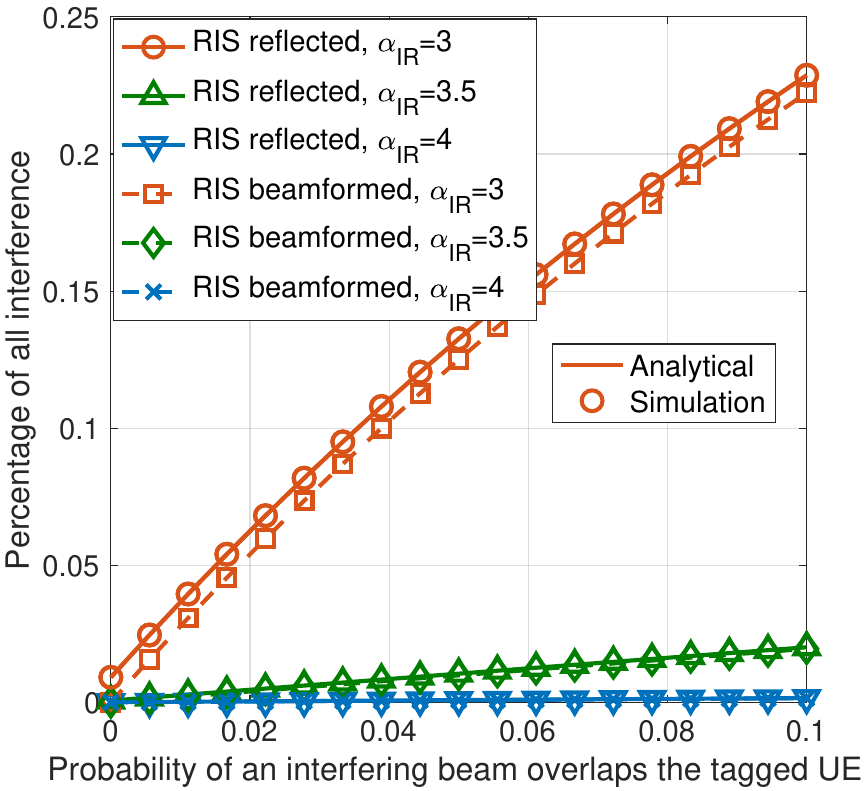}
    \caption{Impact of overlapping probability and cross-cell channel condition on RIS reflected interference percentage}
    \label{fig:RIS_percentage}
\end{figure}
Deploying RIS clusters around all BSs can introduce inter-cell interference. 
Here, the intra-cell interference is neglected since we assume that the BSs can coordinate the frequency resource to the UEs within their own cell.
When RISs reflect the intended signals using narrow beams to the associated UEs, they might also reach the non-associated UEs in other cells that are reusing the same frequency band.
The strength of inter-cell interference due to RIS reflection depends on the channel condition across cells and the probability of the reflected beam reaching the typical UE, as discussed in Lemma~\ref{lemma_reflected_interference}.
Such a dependency is quantitatively shown in Fig.~\ref{fig:RIS_percentage}, where the proportion of power of interference caused by RIS reflection is depicted\footnote{The value used to plot the relationship is computed by evaluating the first-order derivative of the Laplace transform at zero and further verified by Monte Carlo simulation.}.
When the inter-cell paths between RISs and the typical UE have nice channel conditions ($\alpha_{I_R}=3$), the reflected interference strength can be significant when a significant proportion of interfering beams are focused toward the typical UE. 
However, this case is rare in practice since the overlap probability should be low when the interfering beam is narrow and the beams are randomly directed in an arbitrary direction. 
In this analysis, we consider a conservative assumption where the beamwidth is 10 degrees out of 360 degrees, the overlapping probability is only 0.028. 
Thus, as shown in Fig.~\ref{fig:RIS_percentage}, the interference reflected from RISs is insignificant.
Furthermore, in real-world scenarios, inter-cell channels are more likely to be worse than LoS links ($\alpha_{I_R}=\{ 3.5, 4\}$).  
In these cases, the interference caused by RIS reflections is negligible compared to the interference from neighboring BSs, as suggested by the curves in Fig.~\ref{fig:RIS_percentage}. 
Based on the results, we can conclude that the interference reflected by RISs can be neglected when the reflected beam is narrow or the inter-cell channels are worse than LoS links ($\alpha_{I_R}=3.5>\alpha_{\rm LoS}=3$).


\subsubsection{Network constraints: interference-limited or noise-limited}
\begin{figure}
    \centering
\includegraphics[width=0.8\linewidth]{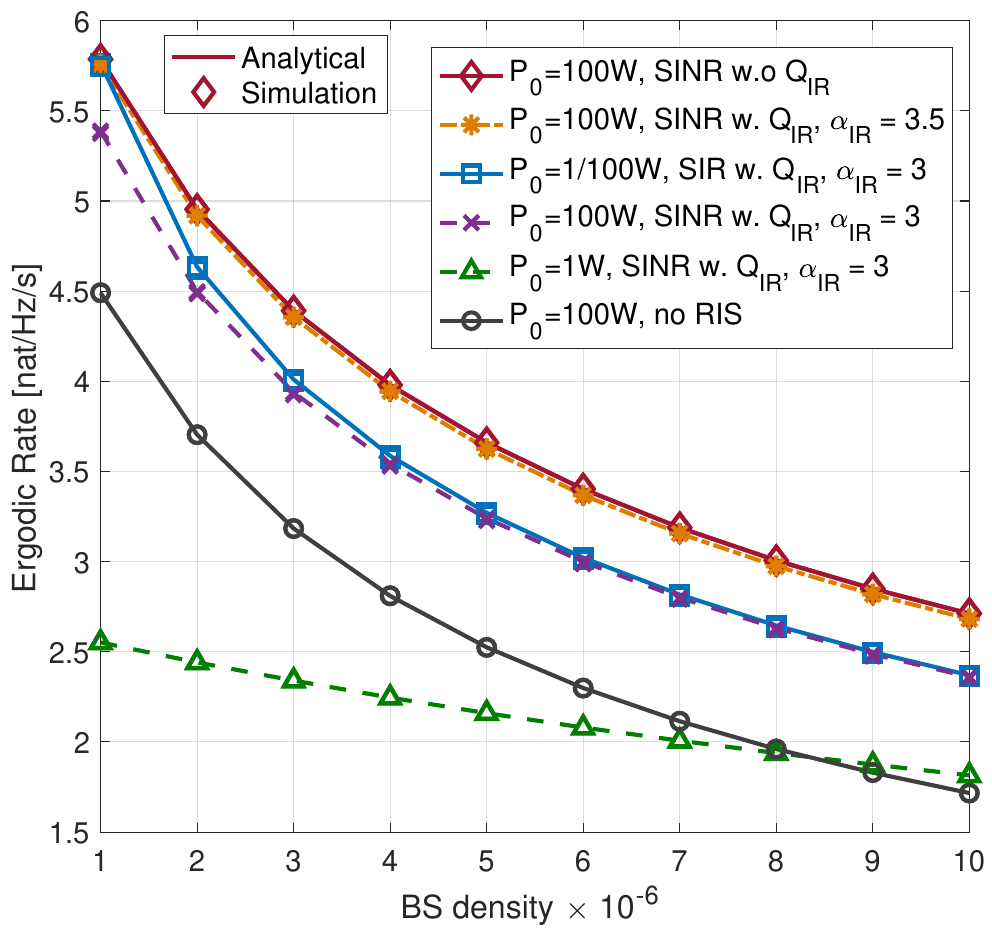}
    \caption{Ergodic rate based on SNR, SIR, SINR, where interference can include the RIS reflected part or not}
    \label{fig:PowerOverall}
\end{figure}

To further analyze the overall impact of interference, we assess the ergodic rate of the typical UE based on either SIR or SINR metrics. We consider scenarios with and without reflected interference, and transmission power of $P_0=100$ Watts additionally, as shown in Fig.~\ref{fig:PowerOverall}.
Specifically, to ensure a conservative estimate (where we intentionally overestimate the reflected interference), we focus on the scenario with the reflected beamwidth of $3.6^{\circ}$ and the pathloss exponent for reflected paths $\alpha_{I_R}=\{3, 3.5 \}$, as detailed in the labels. 

Figure~\ref{fig:PowerOverall} reveals two factors influencing whether a network is interference-limited: BS density and transmission power.
We first take the SIR curve (with blue square-shaped markers) as the reference, since the transmitting power is then canceled in both the numerator and the denominator of the SIR fraction, implying that the curve is valid for both $P_0=1$ Watt and $P_0=100$ Watts.  
We observe that the curves with transmitting power $P_0=100$ Watts always show a similar trend as the SIR curve, implying that a high transmission power renders the network interference-limited as expected.
However, the curve of $P_0=1$ Watt has a different trend but converges to the SIR curve when the BS density is high.
This is because, at low transmission power levels, cellular network performance is primarily limited by noise but when the BS density is high such as $10$/km${}^2$, the interference level becomes comparable to the noise level and the network becomes interference-limited.
Comparing the curve considering RIS reflected interference, i.e., w. $Q_{I_R}$ (that with cross markers),  against the curve w.o. $Q_{I_R}$ (that with diamond markers), we observe that there is a gap between the curves, showing that when the pathloss of the reflected interference path is $\alpha_{I_R} = 3$, the reflected interference cannot be negligible. 
However, when we assume the reflected path is $\alpha_{I_R} = 3.5$, the curve with star markers almost overlaps with that with diamond markers showing that the RISs have a minimal effect on overall interference. 
This further validates the results in the previous simulation that the impact of inter-cell RIS reflected interference can be neglected when the pathloss of the reflected signal is large.
These conditions are easily guaranteed since RIS is typically large, and a large number of elements will create a narrow beam, ensuring a low probability of overlap. 
Additionally, RIS deployment in practice is primarily directed at the associated UEs within their own cell, thus minimizing the possibility of generating interference that affects neighboring cells.
To account for more realistic environment parameters, such as probabilistic blockage, we can use more complex channel models and incorporate them into our framework in future work.

\begin{figure}[htp!]
    \centering
    \includegraphics[width=0.8\linewidth]{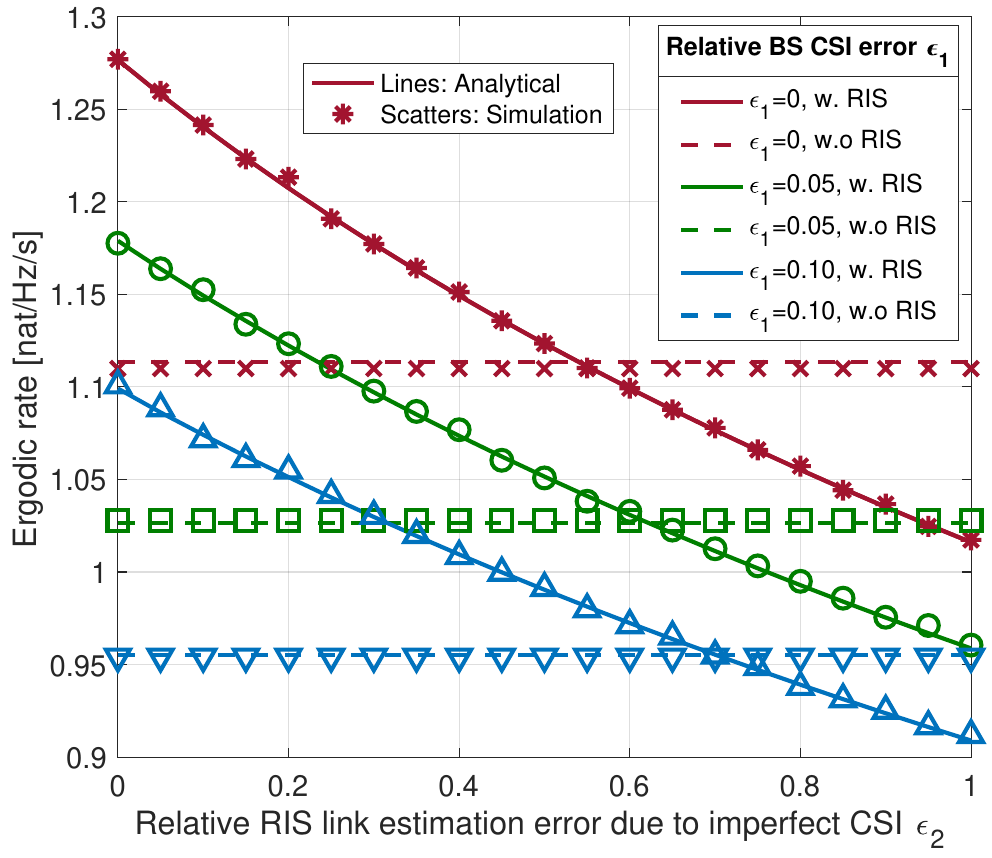}
    \caption{Ergodic rate with and without RIS deployment considering imperfect CSI}
    \label{fig:imperfect_CSI}
\end{figure}

\subsubsection{Imperfect CSI}\label{subsec:imCSI}
Using the MCP model, we analyze the ergodic rate of the RIS-assisted system under imperfect CSI, comparing it to a system without RIS deployment. 
In Fig.~\ref{fig:imperfect_CSI}, we vary the relative estimation error in RIS links (i.e., imperfect CSI) for three different levels of error in the BS link channel estimation. 
This setup reflects the impact of the maturity of BS link channel estimation techniques compared to the developing RIS link estimation and configuration.
We observe that RIS deployment improves system performance with low channel estimation error. 
However, a high channel estimation error in RIS links can even degrade system performance, making a no-RIS deployment preferable. 
This is because we adopt a conservative coding rate to mitigate the uncertainty in the RIS reflected link.  
The results indicate that, until the technology matures, overly conservative coding rate selection, driven by too many factors, can be detrimental to system performance.
Furthermore, a general trend shows that a higher BS link channel estimation error reduces the overall ergodic rate. However, it shifts the crossover point between RIS deployment and no RIS deployment to the right. 
This shift suggests that the BS link channel estimation quality constrains the acceptable RIS link channel estimation error from the system designer's perspective.

\subsubsection{Network configuration}
\begin{figure}
    \centering
    \includegraphics[width=0.8\linewidth]{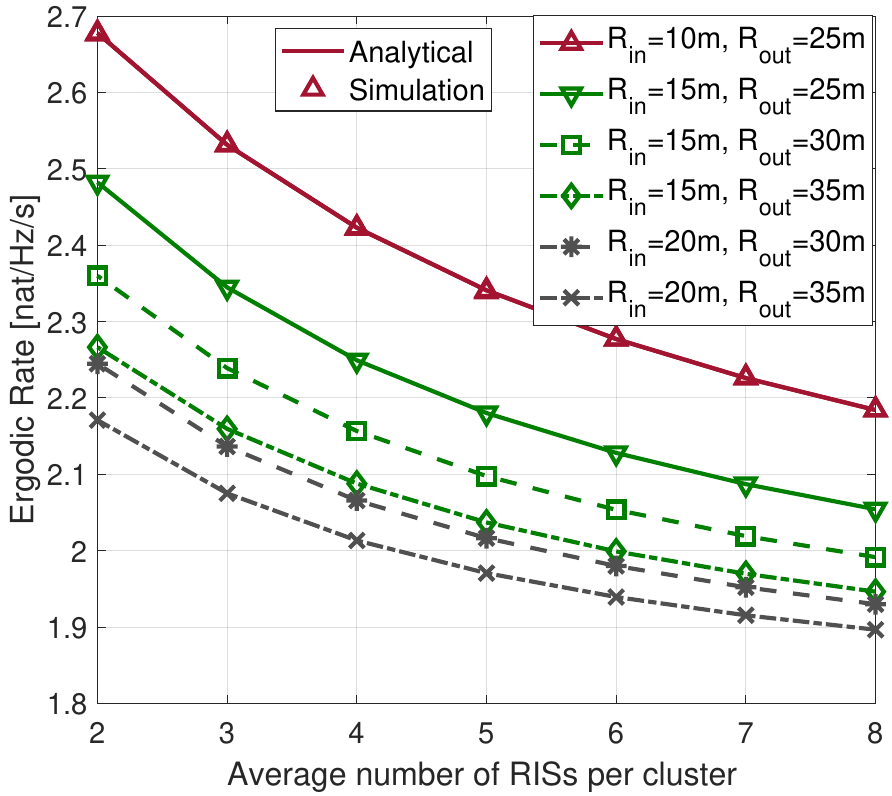}
    \caption{Impact of the density of the RISs, when the total RIS resource is constant}
    \label{fig:smallBigRIS}
\end{figure}
In this paragraph, we evaluate the impact of the size of RIS w.r.t. the density of RISs when the total number of RIS elements per cluster is 10000.
Since the RIS resources are shared, the size of the RIS to perform the beamforming to the typical UE is inversely proportional to the density of the RISs.  
Fig.~\ref{fig:smallBigRIS} plots the relationship between the ergodic rate and the density of RISs for serving the UE located at 100m from the associated BS.
We observe the general trend that, in this setting, a higher RIS density leads to a lower ergodic rate, which shows that, at a system level, fewer but bigger RISs can assist the cellular system better than smaller but more numerous RISs. 
This is because the beamforming gain mainly depends on the number of RIS elements of a batch available to form the beam.
The results show that the spatial diversity gain is not as high as the beamforming gain. 
However, reflected beams might be blocked in a practical network. 
When considering potential blockage, a scenario with multiple, weaker reflected beams becomes more favorable for partial success than a single, powerful beam. Hence, the diversity gain can outperform the beamforming gain.
 
Finally, we investigate the performance improvement w.r.t. different cluster radii. 
When the inner radius is set, the relative gains always decrease when the outer radius increases, implying that a smaller outer radius is favorable.
On the other hand, comparing the line sets for the inner radius, the curves show that a smaller inner radius always outperforms a larger radius, suggesting the superiority of a smaller inner radius. 
Nevertheless, smaller cluster radii suggest that RISs should always be deployed close to the BS. 
This conclusion is valid for the current pathloss model without considering blockage. 
It is important to consider that, in real-world scenarios, obstacles blocking the direct link from the BS might also block the link associated with the RIS that is close to the BS. 
This shared blockage could limit the effectiveness of this particular RIS.
Future research on blockage modeling should account for this possibility to refine the overall performance assessment.

\subsection{Extensions and Variants:}
\begin{figure}
    \centering
    \includegraphics[width=0.8\linewidth]{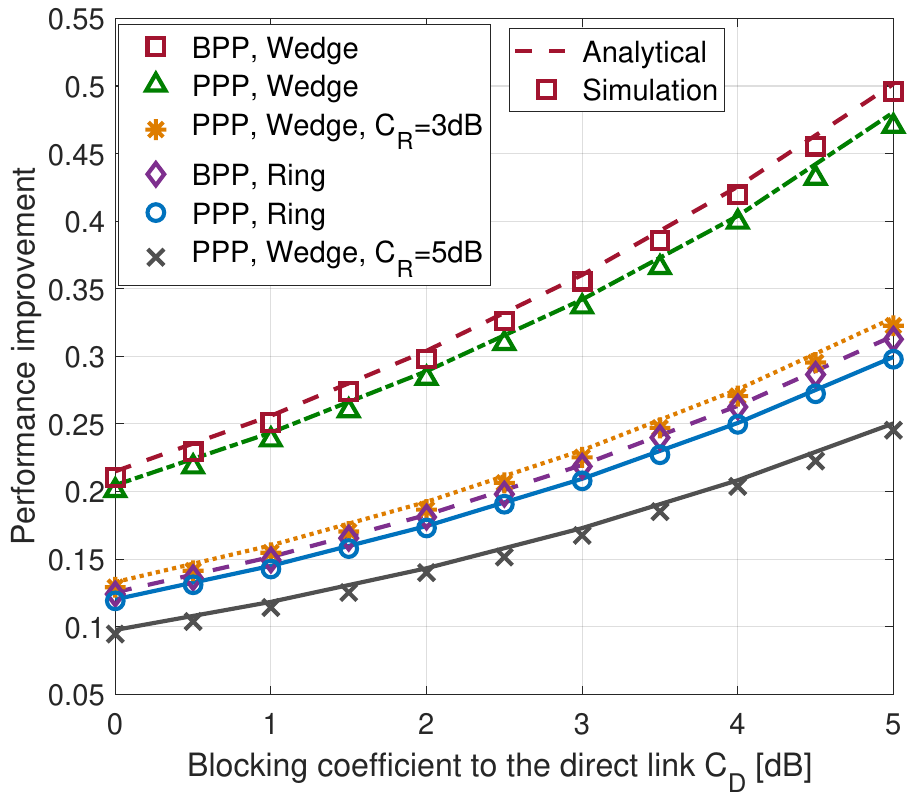}
    \caption{Relative ergodic rate improvement in function of the blocking coefficient $C_{D}$.}
    \label{fig:coverage_hole}
\end{figure}

This subsection demonstrates the versatility of the framework introduced in Section~\ref{section:extensions}.  
We evaluate the ergodic rate w.r.t. multiple variants: impact of blockage, wedge-shaped RIS deployment, and fixed RIS number. 
In the following, we focus on the impact of the spatial deployment of RISs in the associated cell and neglect the interference reflected from RISs in other cells, since we do not specify the deployment of RISs in their cell. 
This simplification is reasonable in the case where the channel condition and the small beamwidth allow one to neglect the reflected interference from other cells, as shown in the conclusion about the reflected interference in the MCP model. 
We set the BS density to $\lambda_{\rm BS}=4/$km${}^2$. 
The center of the coverage hole is located $r=80$m from the typical BS. 
This parameter ensures that the RISs are deployed within a distance less than half of the average distance between BSs.
We set the inner radius to 25 meters and the outer radius to 35 meters, surrounding the coverage hole.
RISs can be deployed over the whole ring, or refined in a wedge-shaped area of $90^\circ$, whose direction aligns with the center of the coverage hole. 
In addition to the PPP modeling, we also simulate the BPP to model the case of the fixed number of RISs with $N_{\rm RIS}=4$. 
Similar to the previous section, each RIS has $M=3000$ elements to serve $5$ UEs in this cell, i.e., each RIS serves each UE using $M_o=600$ elements to form a narrow reflecting beam.
Fig.~\ref{fig:coverage_hole} plots the relative gain when the direct link is affected by a constant blocking penalty $C_{D}$ ranging from 0 to 5 dB and $C_{R} \in \{3, 5\}$ dB. 

In Fig.~\ref{fig:coverage_hole}, the general trend shows that the improvement of the ergodic rate caused by the RISs will increase when the blocking coefficient attached to the direct link is larger, implying the importance of RISs for weak direct links.
In other words, RISs can provide a significant gain in terms of ergodic rate when the direct link is severely blocked. 
Moreover, in the absence of blocking the reflected link, the use of a wedge-shaped area will increase the improvement resulting from RISs. 
This suggests that deploying RISs more strategically will introduce extra benefits. 
However, it is possible that the same obstacle can block both the direct link and the reflected links in this wedge-shaped layout. 
Introducing the blockage penalty $C_{R}=3$dB in the reflected links for the case of the wedge-shaped area can diminish the improvement.
When $C_{R}=5$dB, the improvement is even worse than deploying RISs over the ring. 
In addition, we observe that the improvement provided by a fixed number of RISs (modeled by BPP) is higher than that of a random number of RISs (modeled by PPP). This phenomenon shows that a fixed number of RISs should be preferred.  

In practice, the deployment strategy depends on the scenarios and the corresponding statistics of the network.
Real networks will impose constraints on the deployment area, penalties on the direct link, as well as the number of deployed RISs in a specific area.
We can achieve system-level performance improvements by optimizing the set of parameters based on real-world information using the proposed framework.




\section{Future Works}\label{section:future_works}
This work focuses on characterizing the aggregated signal from randomly located RISs within a defined area. 
We outline future research directions for incorporating more realistic models:
\begin{itemize}
    \item An analytical characterization of the dependence of the number of RISs on cell size is challenging due to the fact that the distribution of 2D Voronoi cell size is not known. However, further investigation in this direction would be useful.
    \item To accurately reflect advancements in RIS technology and configuration algorithms, more sophisticated link-level reflection models are needed. 
    \item More complex blockage models can be built upon this framework by incorporating various statistical descriptions that account for factors like propagation distance, environment type, and carrier frequency. 
    \item The techno-economic analysis of RIS technology can be performed by analyzing how system performance varies with BS and RIS density, considering their deployment costs. This is investigated in a companion paper~\cite{sun2025stochastic}. 
\end{itemize}
Based on this framework, these additional models will have to be investigated to get a more comprehensive understanding of the role of RIS in future cellular networks.
    
\section{Conclusion}\label{section:conclusion}

In this work, we provide a novel system-level analytical framework to evaluate the performance of a RIS-assisted network, where both BSs and RISs are modeled by spatial stochastic processes.
This framework associates a set of RISs with a BS cell. It allows the parameters, such as the deployment area, the number of RISs, and the size of RISs, to be tuned based on deployment strategies.
We focus on a BS-centric deployment modeled by MCP, for which we present the methodology for deriving the system metrics, i.e., the spectrum efficiency, based on the system parameters. 
Numerical results from the MCP model allow one to quantitatively investigate the role of the RIS reflections at a system level and understand the impact of the parameters for RIS deployment.
We then extend the BS-centric MCP model to a UE-centric variant model, which accounts for the alleviation of coverage holes. 
Numerical results on this variant further provide insights into the design of RIS deployment in complex cellular networks. 
To sum up, the BS- and UE-centric models jointly demonstrate the capability and generality of the proposed framework in analyzing RIS-assisted networks.

\appendix
{
\subsection{The small scale fading for the reflected link $\zeta = \rho_{R_1}\rho_{R_2}$}\label{app_sub:zeta}
We discuss the random variable $\zeta\in \mathbb{C}$ that represents the composite small-scale fading of the signal that is reflected by a RIS element\cite{qian2020beamforming}. 
Since we assume that the fading of the link from the BS to the RIS is independent of that of the link from the RIS to the UE, $\zeta$ is the product of two independent small-scale fading variables $\rho_{R}$, given by
\begin{equation}\label{eq:def_zeta}
\zeta = \rho_{R_1}\rho_{
    R_2}  = |\rho_{R_1} | \cdot  | \rho_{R_2}| \cdot e^{\imath(\theta_{R_1}+ \theta_{R_2})}\in \mathbb{C}.  
\end{equation}
According to the fading assumption, the signal reflected by one RIS element experiences two consecutive Rician fades, in the case where the links from the BS to the RISs and from the RISs to the UE have LoS channel conditions.
Nevertheless, the reflected signal, either arriving at the receiver antenna in-phase when beamformed or out-of-phase when scattered from an RIS, is the superposition of reflection from a large number of elements of that same RIS. 
Since the path loss of the reflected path via the same RIS is of the same order, we will later apply the central limit theorem for the superposed fading so that the first two moments of the $|\zeta|$ are sufficient to characterize the superposed reflected signal. 
Classical results \cite{springer1979algebra} on the mean and variance of the product of independently distributed random variables,
\begin{equation}\label{eq:product_fading}
  \begin{aligned}
    \mathbb{E}[|\rho_{R_1} | \cdot  | \rho_{R_2}|] =& \mathbb{E}[|\rho_{R_1} |]\mathbb{E}[| \rho_{R_2}|], \\
	\mathbb{V}[|\rho_{R_1} | \cdot  | \rho_{R_2}|] =& \mathbb{V}[|\rho_{R_1} |]\mathbb{V}[| \rho_{R_2}|] +  \mathbb{V}[|\rho_{R_1} |]\mathbb{E}[| \rho_{R_2}|]^2 + \\ 
    &\mathbb{V}[| \rho_{R_2}|]\mathbb{E}[|\rho_{R_1} |]^2.
\end{aligned}  
\end{equation}

For example, we assume the Rician factor\footnote{The ratio of the energy of the LoS component to that of the scattered component~\cite{tse2005fundamentals}.} of the Rician fading is one in the simulation, without loss of generality, and that the magnitude of fading for each link is distributed as $|\rho_R| \sim \Big| \sqrt{\frac{1}{2} } + \sqrt{\frac{1}{2}}\mathcal{CN}(0, 1)\Big| $.
We have $\mathbb{E}\big[|\rho_R|\big] = \sqrt{\frac{\pi}{8}} {}_1F_1(-\frac{1}{2}; 1; -1) =0.9065 $ and $\mathbb{V}\big[|\rho_R|\big] = 1- \frac{\pi}{8}{}_1F_1(-\frac{1}{2}; 1; -1)^2 = 0.1783$, where ${}_{1}F_{1}(\cdot , \cdot , \cdot)$ denotes the confluent hypergeometric function\cite{andrews1999special}.
Plugging these values in Eq.~\eqref{eq:product_fading}, we get $\mathbb{E}[| \zeta |] = 0.8217$ and $\mathbb{V}[| \zeta |] = 0.3249$.

}

\subsection{Laplace transform of the aggregated interference from a cell $\mathcal{L}_{Q_{c}(x)}(s)$}\label{app:interference}

The interference power from an arbitrary cell (indexed by $i$) is given by $Q_{c_{i}} = \gamma_{I_{D_i}}P_0g(\|\mathbf{x}_i\|) + \sum_{j\in\mathcal{J}_{i}}\gamma_{I_{R_{i,j}}}P_0 G(\mathbf{x}_i, \mathbf{y}_{i,j})$. 
Let $x_i = \|\mathbf{x}_i\|$ denote the distance between the BS of that cell to the typical UE, we have the Laplace transform of the scaled interference $TQ_{c_{i}}$ of this cell $\mathcal{L}_{Q_{c_{i}}}(s)$
\begin{equation}\label{eq:laplace_cluster}
\begin{aligned}[b]
    & \mathbb{E}_{\gamma_{I_{D_i}}, \gamma_{I_{R_{i,j}}}, \phi_{i}}\Big[e^{ -s P_0\big(\gamma_{I_{D_i}}g(x_i) + \sum_{j\in\mathcal{J}_{i}}\gamma_{I_{R_{i,j}}} G(\mathbf{x}_i, \mathbf{y}_{i,j}\big)}  \Big] \\
    =& \mathbb{E}_{\gamma_{I_{D_i}}}\!\Big[\! e^{\!-sP_0 \gamma_{I_{D_i}}\!g(x_i)}\!\Big]\! \mathbb{E}_{ \gamma_{\!I_{R_{i,\!j}}}, \phi_{i}}\!\Big[\!e^{\!-\!  \sum_{j\in\mathcal{J}_{i}}\!sTP \! \gamma_{I_{R_{i,j}}} \!G(\mathbf{x}_i, \mathbf{y}_{i,j}) } \!\Big]\\
    \stackrel{(a)}{=}&\mathcal{L}_{\gamma_{I_D}\!}\!\big(\!sP_0g(x_i)\!\big)\! \mathbb{E}_{ \gamma_{I_{R_{i,j\!}}}, \phi_{i}}\!\Big[\!e^{\!- \! \sum_{j\in\mathcal{J}_{i}\!}\!sT\!P_0 \!\gamma_{I_{R_{i,j}}} \!\!G(\mathbf{x}_i, \mathbf{y}_{i,j}\!)}\! \Big|O_{\rm beam}\Big]\\
    & \mathbb{E}_{ \gamma_{I_{R_{i,j}}}, \phi_{i}}\Big[e^{-  \sum_{j\in\mathcal{J}_{i}}sTP_0 \gamma_{I_{R_{i,j}}} G(\mathbf{x}_i, \mathbf{y}_{i,j})} \Big|\bar{O}_{\rm beam}\Big]\\
    \stackrel{(b)}{=}&\frac{1}{1\!+\!sP_0g(\!x_i\!)}e^{\!-\!\frac{\vartheta_{\rm beam}}{2\pi} \! \lambda_{\rm RIS} \!\int_{0}^{2\pi}\!\!\int_{R_{\rm in}}^{R_{\rm out}} \!\big(\!1\!-\!\mathcal{L}_{\gamma_{S_R}}(sP_0\mathcal{G}(x_i, y, \psi))\!\big)\! {\rm d}y{\rm d}\psi}\\
    &e^{-\big(1-\frac{\vartheta_{\rm beam}}{2\pi}\big) \lambda_{\rm RIS}  \int_{0}^{2\pi}\int_{R_{\rm in}}^{R_{\rm out}} \big(1-\frac{1}{1+s(M-M_o)P_0\mathcal{G}(x_i, y, \psi)}\big) {\rm d}y{\rm d}\psi}.
\end{aligned}
\end{equation}
where $\mathcal{L}_{\gamma_{S_R}}(s)$ is given in Lemma~\ref{lemma_gamma_sr}. Here in $(a)$, we use the fact that the reflected interference from any given RIS arrives at the typical UE in either the beamformed part or the scattered part. 
If the typical UE is within the interference beam, we denote the event as $O_{\rm beam}$. Otherwise, the event of experiencing the scattered interference is denoted as $\bar{O}_{\rm beam}$. They are modeled by two thinning Poisson point processes with the density $\frac{\vartheta_{\rm beam}}{2\pi} \lambda_{\rm RIS}$ and $(1-\frac{\vartheta_{\rm beam}}{2\pi}) \lambda_{\rm RIS}$, respectively, where $\frac{\vartheta_{\rm beam}}{2\pi}$ is the fraction of beam over the total angular domain $2\pi$. $(b)$
follows from the fact that $\phi_i$ are Poisson point process and $\gamma_{I_{R_i}}$ are i.i.d. fading, thus we can apply the PGFL of the PPP.

\subsection{Region of convergence for the Laplace transform $\mathcal{B}_{\Upsilon}(s)$}\label{app:convergence}
The proposed analytical method requires the existence of the bilateral Laplace transform, $\mathcal{B}_{\Upsilon}(s)$. This necessitates that the argument of the Laplace functional $s = \frac{1}{P_0g(r)}$ be defined within its region of convergence.

The bilateral Laplace transform, $\mathcal{B}_{\Upsilon}(s) = \int_{-\infty}^{\infty}f(t)e^{-st}{\rm d}t$, converges in a vertical strip of the complex $s$-plane, defined by $s_a<\Re(s)<s_b$, where $-\infty \leq s_a < s_b \leq \infty$\cite[Chapter~VI]{widder2015laplace}.
Recall that we define $\Upsilon = T\big(Q_I(r)+\sigma_{w}^2 {+\epsilon_1 g(r)+\epsilon_2\mathbb{E}[Q_{S_R}(r)]}\big)-Q_{S_R}(r)$ (Eq.~\eqref{eq:definition}).
The non-negativity of the term, $Q_I(r)+\sigma_{w}^2 {+\epsilon_1 g(r)+\epsilon_2\mathbb{E}[Q_{S_R}(r)]}$, and  the non-positivity of the term, $-Q_{S_R}(r)$ determine the region of convergence for the Laplace transform $\mathcal{B}_{\Upsilon}(s)$. 
Specifically, the Laplace transform of non-negative random variable $X$, given by $\int_0^{\infty}e^{-sx}f(x){\rm d}x$, converge in the right half-plane $\Re{(s)}>s_a$. Conversely, the Laplace transform of the non-positive term, $-Q_{S_R}(r)$, converges in the left half plane $\Re{(s)}<s_b$.

Due to the characteristic function of any PDF exists, we have $s_a<0<s_b$.
In addition, the argument-to-evaluate of the Laplace transform for computing the coverage probability is $s = \frac{1}{ P_0g(r)}>0$, as given in Eq.~(27). 
Therefore, we focus on discussing $s_b$ imposed by the aggregated reflection signal since the region of interest is $0<\Re(s)<s_b$. 
In other words, the value of $s_a$ does not impact the convergence behavior. 
To identify the constraint on $s_b$, we we examine the convergence of the Laplace transform given in Lemma~\ref{statement1}. 
By Eq.~\eqref{eq:reflecting_signal}, the convergence indeed depends on whether $\mathcal{L}_{\gamma_{S_R} }\big(-s P_0\mathcal{G}(r, y, \psi)\big)$ converges for all $r, y, \psi$. 
Substituting $s=\frac{1}{P_0g(r)}$ requires that the argument $ s' = -\frac{\mathcal{G}(r, y, \psi)}{ g(r)}$ falls within the region of convergence for $\gamma_{S_R}$.
The region of convergence for the Laplace transform of a standard non-central $\chi^2$ distribution is $s'>-\frac{1}{2}$ and that of an exponential distribution is $s'>-1$.
Inserting variance $\sigma_{\Re}^2$ defined in Lemma~\ref{lemma_gamma_sr}, we obtain the implicit constraint that to guarantee the convergence
\begin{equation}\label{LT_constraint}
    \frac{(M+M_o)\mathbb{V}[|\zeta|] \mathcal{G}(r, y, \psi)}{2 g(r)} < \frac{1}{2}.
\end{equation}
When the parameters of the network satisfy the constraints in Eq.~\eqref{LT_constraint}, we could apply the analytical method to assess the network performance.
All subsequent analyses must ensure parameters remain within this defined region.

\subsection{Expressing $\mathcal{B}_{\Upsilon^+}(s)$ in function of $\mathcal{B}_\Upsilon(s)$}\label{app:max_variable}

For a real-valued variable $x\in \mathbb{R}$, define $x^+ = \max \{x, 0\}$,  and denote the sign function $${\rm sign}(x) = \left\{ 
\begin{aligned}
&-1, \quad &x<0, \\ 
&1, \quad &x\geq 0.
\end{aligned}
\right.$$
Then, the following identity~\cite{pinelis2015characteristic} holds based on manipulations of $e^{-sx}$:  $$e^{-sx^+} - e^{-sx} = 
\frac{1}{2}\big(e^{-s\cdot 0} - e^{-sx}\big)\big(1-{\rm sign}(x)\big),$$ which gives 
\begin{equation}\label{eq:app_}
\begin{aligned}
 e^{-sx^+} =& e^{-sx} + \frac{1}{2}\Big(1 - e^{-sx}\Big)\Big(1-{\rm sign}(x)\Big)\\
 =& \frac{1}{2}\Big(1+e^{-sx} - \big(1-e^{-sx}\big) \cdot {\rm sign}(x) \Big).
\end{aligned}
\end{equation}
Using the identity $\int_{-\infty}^{\infty}\frac{\sin(ux)}{u}{\rm d}u = \pi {\rm sign}(x)$, it follows that  
\begin{equation}\label{eq:exp_positive}
\begin{aligned}
    e^{-sx^+}  = & \frac{1}{2}\Big(1+e^{-sx} - \big(1-e^{-sx}\big)\frac{1}{\pi}\int_{-\infty}^{\infty}\frac{\sin(ux)}{u}{\rm d}u \Big) \\
     \stackrel{(a)}{=} & \frac{1}{2}\Big(1+e^{-sx} - \big(1-e^{-sx}\big)\frac{1}{\pi}\int_{-\infty}^{\infty}\frac{e^{\imath ux}-e^{-\imath ux}}{2\imath u}{\rm d}u \Big)  \\
     = & \frac{1}{2}\bigg(1+e^{-sx} + \frac{1}{2\pi \imath}\int_{-\infty}^{\infty} \Big(e^{-sx+\imath ux}-e^{-sx-\imath ux} \\
    &- e^{\imath ux} + e^{-\imath ux} \Big)\frac{{\rm d}u}{u} \bigg) \\
     \stackrel{(b)}{=}&  \frac{1}{2}\bigg(1+e^{-sx} + \frac{1}{\pi \imath}\int_{-\infty}^{\infty}\Big( e^{-sx+\imath ux} - e^{\imath ux} \Big)\frac{{\rm d}u}{u} \bigg), 
\end{aligned}
\end{equation}
where $(a)$ follows from the Euler formula $\sin(x) = \frac{e^{\imath x}-e^{-\imath x}}{2\imath}$, and $(b)$ follows from the fact that the integral defined in the principal value sense is an odd symmetric function. 

Applying the expectation operator $\mathbb{E}[\cdot]$  over both side of Eq.~\eqref{eq:exp_positive}, we have
\begin{equation}\label{eq:positive_bilateral}
\begin{aligned}
&\mathcal{B}_{\Upsilon^+}(s)=\mathbb{E}\Big[ e^{-s\Upsilon^+}\Big] = \int_{-\infty}^{\infty}e^{-sx^+} f_{\Upsilon}(x) {\rm d}x 
\\ =&\int_{-\infty}^{\infty}\bigg[ \frac{1}{2\pi \imath} \int_{-\infty}^{\infty}( e^{-sx+\imath u x} - e^{\imath u x})\frac{{\rm d}u}{u} + \frac{1+e^{-sx}}{2}\bigg]f_{\Upsilon}(x){\rm d}x\\
 =&  \int_{-\infty}^{\infty}\Big( \int_{-\infty}^{\infty} e^{-sx+\imath u x}   f_{\Upsilon}(x){\rm d}x - \int_{-\infty}^{\infty} e^{\imath u x} f_{\Upsilon}(x){\rm d}x \Big) \frac{{\rm d}u}{2\pi\imath u} \\
&+ \frac{1}{2}\bigg(  1 + \int_{-\infty}^{\infty} e^{-sx} f_{\Upsilon}(x){\rm d}x  \bigg) \\
 =&  \frac{1}{2 \pi \imath}\int_{-\infty}^{\infty} \Big( \mathcal{B}_{\Upsilon}(s-\imath u) - 
 \mathcal{B}_{\Upsilon}(-\imath u) \Big) \frac{{\rm d}u}{u} + \frac{1}{2}\Big( 1+ \mathcal{B}_{\Upsilon}(s) \Big), 
\end{aligned}
\end{equation}
when $s$ is in the region of convergence, which is discussed in Appendix~\ref{app:convergence}.

\bibliographystyle{ieeetr}
\bibliography{reference}


\end{document}